\documentclass[twocolumn]{aastex631}

\newcommand{\src}{1E\,1841$-$045}

\newcommand{\srcone}{4U\,0142+61}
\newcommand{\srctwo}{1RXS\,J170849.0--400910}
\newcommand{\srctwosh}{1RXS\,J1708}
\newcommand{\ixpe}{{\rm IXPE}}
\newcommand{\swift}{{\rm Swift}}
\newcommand{\nicer}{{\rm NICER}}

\newcommand{\xmm}{{\rm XMM-Newton}}
\newcommand{\cha}{{\rm Chandra}}
\newcommand{\nustar}{{\rm NuSTAR}}

\newcommand{\nh}{N_\mathrm{H}}
\newcommand{\flux}{\mathrm{erg\,cm}^{-2}\mathrm{\,s}^{-1}}
\newcommand{\lum}{\mathrm{erg}\mathrm{\,s}^{-1}}
\def \radec {$\mathrm{R.A.=18^h41^m19.34^s}$, $\mathrm{Dec.=-04^\circ56'11.16''}$}

\usepackage{graphicx}	
\usepackage{amsmath}	
\usepackage{hyperref}
\usepackage{footmisc} 
\usepackage{multirow}
\usepackage{booktabs}

\begin{document}

\title{IXPE detection of highly polarized X-rays from the magnetar \src}

\correspondingauthor{Michela Rigoselli}
\email{michela.rigoselli@inaf.it}

\author[0000-0001-6641-5450]{Michela Rigoselli} 
\affiliation{INAF, Osservatorio Astronomico di Brera, via Brera 28, I-20121 Milano, Italy}
\affiliation{INAF, Istituto di Astrofisica Spaziale e Fisica Cosmica di Milano, via Corti 12, I-20133 Milano, Italy}

\author[0000-0002-1768-618X]{Roberto Taverna}
\affiliation{Dipartimento di Fisica e Astronomia, Universit\`{a} degli Studi di Padova, Via Marzolo 8, I-35131 Padova, Italy}

\author[0000-0003-3259-7801]{Sandro Mereghetti}
\affiliation{INAF, Istituto di Astrofisica Spaziale e Fisica Cosmica di Milano, via Corti 12, I-20133 Milano, Italy}

\author[0000-0003-3977-8760]{Roberto Turolla}
\affiliation{Dipartimento di Fisica e Astronomia, Universit\`{a} degli Studi di Padova, Via Marzolo 8, I-35131 Padova, Italy}
\affiliation{Mullard Space Science Laboratory, University College London, Holmbury St Mary, Dorking, Surrey RH5 6NT, UK}

\author[0000-0001-5480-6438]{Gian Luca Israel}
\affiliation{INAF, Osservatorio Astronomico di Roma, via Frascati 33, I-00078 Monteporzio Catone, Italy}

\author[0000-0001-5326-880X]{Silvia Zane}
\affiliation{Mullard Space Science Laboratory, University College London, Holmbury St Mary, Dorking, Surrey RH5 6NT, UK}

\author[0009-0001-4644-194X]{Lorenzo Marra}
\affiliation{Dipartimento di Fisica e Astronomia, Universit\`{a} degli Studi di Padova, Via Marzolo 8, I-35131 Padova, Italy}

\author[0000-0003-3331-3794]{Fabio Muleri}
\affiliation{INAF, Istituto di Astrofisica e Planetologia Spaziali, Via del Fosso del Cavaliere 100, I-00133 Roma, Italy}

\author[0000-0001-8785-5922]{Alice Borghese}
\affiliation{European Space Agency (ESA), European Space Astronomy Centre (ESAC), Camino Bajo del Castillo s/n, 28692 Villanueva de la Ca\~{n}ada, Madrid, Spain}

\author[0000-0001-7611-1581]{Francesco Coti Zelati}
\affiliation{Institute of Space Sciences (ICE, CSIC), Campus UAB, Carrer de Can Magrans s/n, E-08193 Barcelona, Spain}
\affiliation{Institut d'Estudis Espacials de Catalunya (IEEC), 08860 Castelldefels (Barcelona), Spain}

\author[0000-0001-5438-0908]{Davide De Grandis}
\affiliation{Institute of Space Sciences (ICE, CSIC), Campus UAB, Carrer de Can Magrans s/n, E-08193 Barcelona, Spain}
\affiliation{Institut d'Estudis Espacials de Catalunya (IEEC), 08860 Castelldefels (Barcelona), Spain}

\author[0000-0001-8688-9784]{Matteo Imbrogno}
\affiliation{Dipartimento di Fisica, Universit\`{a} degli Studi di Roma Tor Vergata, Via della Ricerca Scientifica 1, I-00133, Rome, Italy}
\affiliation{INAF, Osservatorio Astronomico di Roma, via Frascati 33, I-00078 Monteporzio Catone, Italy}

\author[0000-0002-5004-3573]{Ruth M.E. Kelly}
\affiliation{Mullard Space Science Laboratory, University College London, Holmbury St Mary, Dorking, Surrey RH5 6NT, UK}

\author[0000-0003-4849-5092]{Paolo Esposito}
\affiliation{Scuola Universitaria Superiore IUSS Pavia, Palazzo del Broletto, piazza della Vittoria 15, I-27100 Pavia, Italy}

\author[0000-0003-2177-6388]{Nanda Rea}
\affiliation{Institute of Space Sciences (ICE, CSIC), Campus UAB, Carrer de Can Magrans s/n, E-08193 Barcelona, Spain}
\affiliation{Institut d'Estudis Espacials de Catalunya (IEEC), 08860 Castelldefels (Barcelona), Spain}



\begin{abstract}
The Imaging X-ray Polarimetry Explorer (\ixpe) observed for the first time highly polarized X-ray emission from the magnetar \src, targeted after a burst-active phase in August 2024.
To date, \ixpe\ has observed four other magnetars during quiescent periods, highlighting substantially different polarization properties.
\src\ exhibits a high, energy-dependent polarization degree, which increases monotonically from $\approx\!15\%$ at $2$--$3\,\mathrm{keV}$ up to $\approx\!55\%$ at $5.5$--$8\,\mathrm{keV}$, while the polarization angle, aligned with the celestial North, remains fairly constant.
The broadband spectrum ($2$--$79\,\mathrm{keV}$) obtained by combining simultaneous \ixpe\ and \nustar\ data is well modeled by a blackbody and two power-law components. The unabsorbed $2$--$8\,\mathrm{keV}$ flux ($\approx2 \times 10^{-11}\,\flux$) is about $10\%$ higher than that obtained from archival \xmm\ and \nustar\ observations.
The polarization of the soft, thermal component does not exceed $\approx\!25\%$, and may be produced by a condensed surface or a bombarded atmosphere.
The intermediate power law is polarized at around $30\%$, consistent with predictions for resonant Compton scattering in the star magnetosphere; while, 
the hard power law exhibits a polarization degree exceeding $65\%$, pointing to a synchrotron/curvature origin.

\end{abstract}



\section{Introduction} \label{sec:intro}

Magnetars are ultra-magnetized neutron stars whose high-energy emission is mainly powered by their own magnetic energy \citep{td92,td93}. The presence of magnetic fields up to three orders of magnitude stronger than those typically found in ordinary young neutron stars is at the basis of their enhanced  X-ray emission and extreme variability phenomena on different timescales, ranging from short (tens of ms) bursts to months- or even years-long outbursts \citep[see, e.g.,][and references therein]{turolla+15,kaspi+belo17,2021ASSL..461...97E}.

Such a strong magnetic field ($B\sim 10^{13}$--$10^{15}\,\mathrm G$) makes magnetars extremely interesting targets for polarimetric observations in the X-ray range, which have been recently made possible by the NASA-ASI Imaging X-ray Polarimetry Explorer \cite[\ixpe;][]{Weisskopf2022}. 
Radiation in a strongly magnetized medium propagates in the ordinary (O) and extraordinary (X) normal modes \citep{1974JETP...38..903G}.
The opacity for X-mode photons is strongly suppressed, below the electron cyclotron energy, relative to that of the O-mode ones \cite[see][and references therein]{2006RPPh...69.2631H}. The X-ray emission from magnetar sources is, then, expected to be substantially polarized \citep[up to $\approx\!80\%;$][]{fernandez2011,taverna+14,taverna+20,Caiazzo2022}.

The \ixpe\ satellite observed four magnetars during the first two years of operations.
Polarization was clearly measured in the $2$--$8\,\mathrm{keV}$ band for  \srcone\ \citep{taverna+22}, \srctwo\ \citep[hereafter \srctwosh\ for short;][]{zane+23} and 1E 2259$+$586 \citep{2024MNRAS.52712219H},  while an unexpectedly low flux prevented significant detection in SGR 1806$-$20 \citep{2023ApJ...954...88T}.
In the two brightest sources the polarization properties turned out to strongly depend on energy. In \srctwosh, the degree of polarization increases monotonically from $\approx\!20\%$ to $\approx\!80\%$ at a constant polarization angle making it the most polarized source detected by \ixpe\ so far, while in \srcone\ it is $\approx\!15\%$ at $2$--$3\,\mathrm{keV}$, is consistent with zero around $4$--$5\,\mathrm{keV}$,  where the polarization angle swings by $90^\circ$, and then reaches $\approx\!35\%$ at $6$--$8\,\mathrm{keV}$.
 
All the \ixpe\ observations described above were targeted on bright magnetars during periods of quiescence (see also \citealt{2024Galax..12....6T} for a review of all the results). On the other hand, most magnetars exhibit periods of activity during which they emit frequent short bursts of hard X-rays accompanied by flux and spectral changes and/or anomalies in the timing properties (glitches, spin-down rate variations, pulse profile changes). Since these phenomena are probably associated with magnetospheric reconfigurations, it is interesting to explore the polarimetric properties during or after periods of activity. 

The recent reactivation of the magnetar \src\ offered this possibility.
The source lies at the center of the shell-type, $4'$-diameter supernova remnant (SNR) Kes 73, at an estimated distance of $D\approx8.5\,\mathrm{kpc}$ \cite[][]{2008ApJ...677..292T,2010ApJ...725L.191K,2014ApJS..212....6O}.
Its X-ray luminosity $L_\mathrm X \sim 5\times 10^{35}\ \lum$, largely in excess of the rotational energy loss rate, and the detection of pulsations suggested a magnetar interpretation \citep{1997ApJ...486L.129V}. This was confirmed by the subsequent detection of short bursts at different epochs \cite[][]{2003PASJ...55L..45M,2004ApJ...615..887W}.  
Very recently, after more than 9 years of inactivity \citep{2013ApJ...779..163A,2015GCN.18024....1B}, \src\ emitted a series of hard X-ray bursts, beginning on 2024 August 20th \citep{2024GCN.37211....1S,2024GCN.37222....1S,2024GCN.37234....1F,2024GCN.37240....1G,2024GCN.37297....1S}. Soft X-ray observations of the source, performed by \swift-XRT and \nicer\ \citep{2024ATel16789....1D} about one day after the first burst, revealed an enhancement of $\approx\!25$\% of the $0.5$--$10\,\mathrm{keV}$ flux with respect to the quiescent level.

In this \emph{Letter} we report on an \ixpe\ target of opportunity (ToO) observation of \src\ performed soon after the burst-active phase, along with \nustar\ and archival \cha, \xmm\ and \nustar\ campaigns.  Observations are detailed in \S\ref{sec:observation} and the results of timing, spectral, and polarimetric analyses are presented in \S\ref{sec:results}. Discussion follows in \S\ref{sec:discuss}.
In a companion paper \citep{2024arXiv241216036S} an independent analysis of the same \ixpe\ datasets is presented. The results and theoretical interpretation are largely consistent, except for the discussion on the nature of the soft X-ray component, where the two papers present distinct interpretive elements.

\section{Observations and data analysis} \label{sec:observation}

\setlength{\tabcolsep}{0.6em}
\begin{table*}
\centering \caption{Log of X-ray observations and measured spin periods used in this work}
\label{tab:log}

\begin{tabular}{lcccccc}
\toprule
\midrule
Observatory & Instrument/Mode & ObsID & Start time & End time  & Exposure  & $P$\\[3pt]
  &  &  &  (UTC)     &  (UTC)    & (ks) & (s) \\[3pt]
\midrule
\cha & ACIS/TE &   729 & 2000-07-23 20:55:29 & 2000-07-24 05:46:55 & 29.26 &  --\\
\cha & ACIS/TE &  6732 & 2006-07-30 06:54:37 & 2006-07-30 14:30:26 & 24.86 &  --\\
\cha & ACIS/TE & 16950 & 2015-06-04 20:42:20 & 2015-06-05 05:16:21 & 28.69 &  --\\
\cha & ACIS/TE & 17668 & 2015-07-07 11:21:02 & 2015-07-07 17:32:53 & 20.89 &  --\\
\cha & ACIS/TE & 17692 & 2015-07-08 22:57:29 & 2015-07-09 05:54:24 & 23.26 &  --\\
\cha & ACIS/TE & 17693 & 2015-07-09 22:00:45 & 2015-07-10 04:58:34 & 22.77 &  --\\
\xmm & EPIC/FF & 0783080101 & 2017-03-31 05:54:27 & 2017-04-01 04:41:49 & 71.45 &  11.79693(2) \\
\nustar &  --   & 30001025012 & 2013-09-21 11:26:07 & 2013-09-23 22:51:07 & 100.5 & 11.792407(8)\\
\nustar &  --  & 91001330002 & 2024-08-29 04:21:09 & 2024-08-30 07:11:09 & 50.46 & 11.80646(3)\\
\nustar &  --  & 91001335002 & 2024-09-28 23:36:09 & 2024-09-30 07:36:09 & 55.36 & 11.80654(2)\\
\ixpe     &    --  &  03250499   & 2024-09-28 01:32:39 & 2024-10-10 05:12:53 & 292.5 & 11.80659(1)\\

\bottomrule\\[-5pt]
\end{tabular}

\raggedright
\end{table*}

 \subsection{\ixpe}
\label{subsec:ixpe}

Following reports of bursting activity and rebrightening of \src, an \ixpe\ ToO pointing was requested (PI G.\ Younes). The observation, divided into two segments, started on 2024 September 28 01:32:39 UTC and ended on 2024 October 10 05:12:53 UTC, for a total of $\approx\!290\,\mathrm{ks}$ of on source time for each of the three detector units (DUs). 
 
We retrieved the level 1 and level 2 photon lists from the \ixpe\ archive\footnote{\url{https://heasarc.gsfc.nasa.gov/docs/ixpe/archive/}} 
and performed background rejection \cite[see e.g.][]{2023AJ....165..143D}, excising a $\approx\!100\,\mathrm{s}$ interval, which includes solar flares. Source counts were extracted from a circular region with radius $48''$, centered on the position of \src. We checked that by taking a source extraction radius of $30''$ \citep{Weisskopf2022} the overall counts decrease (especially at higher energies), implying that some source events are lost without a definite improvement of the signal-to-noise (S/N) ratio. We extracted background counts from a concentric annulus with inner and outer radii $r_\mathrm{int}^\mathrm{bkg}=60''$ and $r_\mathrm{ext}^\mathrm{bkg}=240''$ solely for the polarization analysis (detailed in \S\ref{subsubsec:polarizationenergy-resolved} and \S\ref{subsubsec:polarizationphase-resolved}), that is, subtracting the SNR contribution from the source counts as normal background. For the spectral analysis, instead, we selected $r_\mathrm{int}^\mathrm{bkg}=150''$, to include the SNR contribution as well (see \S\ref{subsubsec:specpolanalysis}). 

No significant variation in the \ixpe\ count rates is visible throughout the entire observation, so in the following we analyze the dataset obtained by joining the two segments.
Given that background counts are always subdominant with respect to those of the source within the \ixpe\ band, we opted for an unweighted analysis of the level 2 photon lists, using version 20240701-v013 of the response files, provided in the online calibration database\footnote{\url{https://heasarc.gsfc.nasa.gov/docs/ixpe/caldb}}. We checked that a weighted analysis does not produce significant differences.

\subsection{\cha}
\label{subsec:cha}

\begin{figure}
\begin{center}
\includegraphics[trim={3cm 0cm 1cm 1cm},clip,width=1\columnwidth]{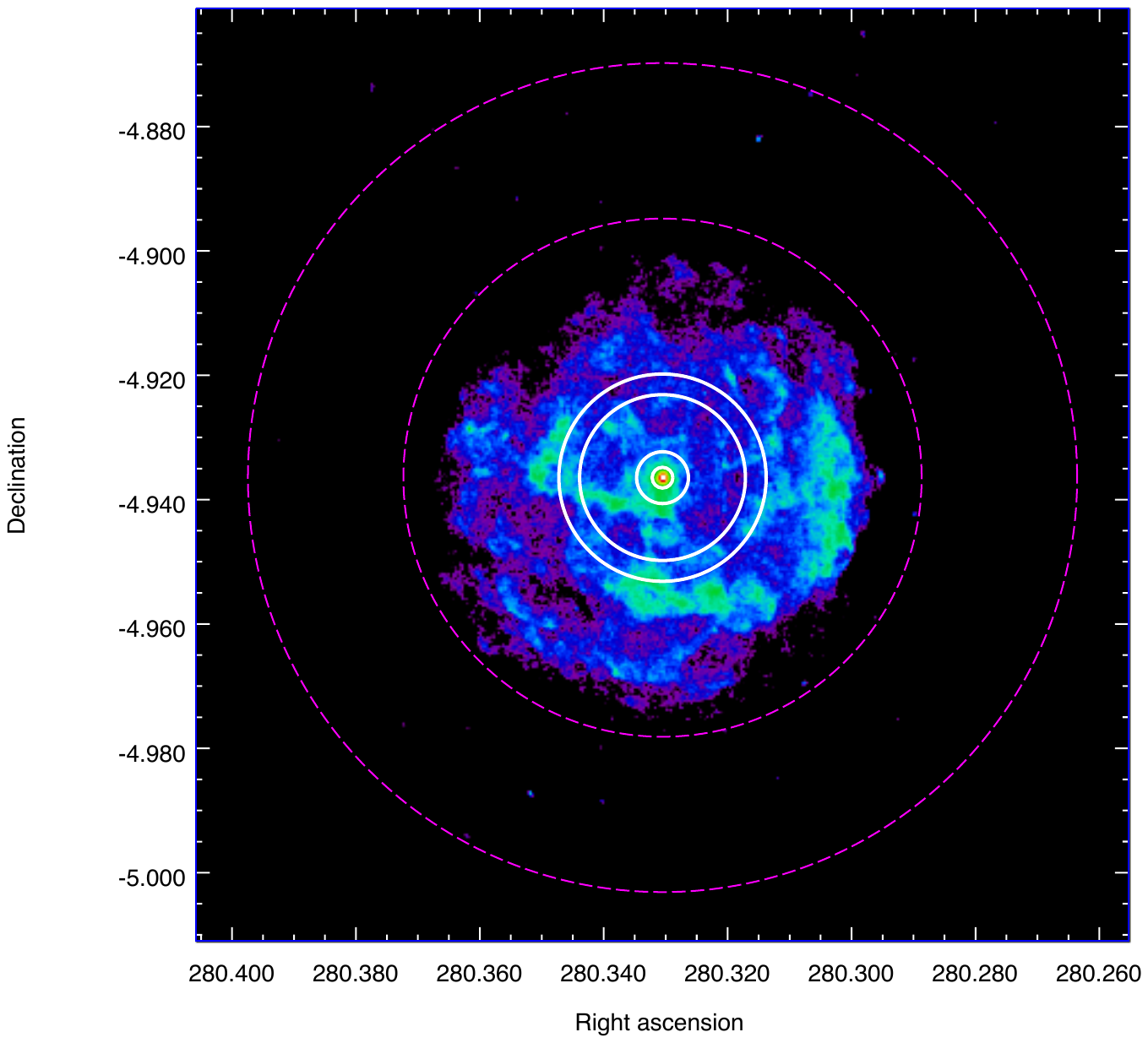}
\caption{\cha\ mosaic of SNR Kes 73. The solid white circles mark the three annular extraction regions for the SNR spectrum, at $r_{\rm int}^{\rm SNR}=6^{\prime\prime}$ (innermost circle, common to all three) and  $r_{\rm ext}^{\rm SNR}=15^{\prime\prime}$, $48^{\prime\prime}$ and  $60^{\prime\prime}$ (outer circles). The dashed magenta circles mark the extraction region used for the background of \xmm, \nustar\ and \ixpe\ data: $r_{\rm int}^{\rm bkg}=150^{\prime\prime}$ and $r_{\rm ext}^{\rm bkg}=240^{\prime\prime}$.
\label{fig:snr}}    
\end{center}
\end{figure}

To estimate the contribution of the Kes 73 SNR to the magnetar spectra obtained with other instruments, we analyzed the archival data obtained with \cha, which, thanks to its unmatched angular resolution, is the best X-ray telescope to disentangle the magnetar and SNR emissions. We used all the ACIS observations in Timed-Exposure (TE) mode listed in Table~\ref{tab:log} (see also \dataset[DOI: 10.25574/cdc.322]{https://doi.org/10.25574/cdc.322}), with the SNR located at the aim point of observation. 
We reprocessed the data with the tool \texttt{chandra\_repro} of the \cha\ Interactive Analysis of Observation software (\texttt{CIAO}; \citealt{2006SPIE.6270E..1VF}), version 4.16, using the calibration database \texttt{CALDB} 4.11.1.
The mosaic of the \cha-ACIS images in the $0.5$--$8\,\mathrm{keV}$ energy band is shown in Figure~\ref{fig:snr}.

For each of the six observations, we extracted three spectra from annuli centered on the magnetar position and with inner radius $r_{\rm int}^{\rm SNR}=6^{\prime\prime}$, to avoid contamination from the magnetar, and outer radii $r_{\rm ext}^{\rm SNR}=15^{\prime\prime}$, $48^{\prime\prime}$, and $60^{\prime\prime}$ (see Figure \ref{fig:snr}). 
These values were chosen to match the extraction regions used for \xmm-EPIC (\S \ref{subsec:xmm}), \ixpe\ (\S \ref{subsec:ixpe}) and \nustar\ (\S \ref{subsec:nustar}), respectively.
We extracted background counts from nearby boxes of size $400^{\prime\prime} \times 150^{\prime\prime}$.

For each set of extraction regions, we simultaneously fit the spectra of the six observations with the sum of two non-equilibrium collisional ionization plasma models, absorbed by the interstellar medium (\textsc{tbabs*(vpshock+vpshock)}, see also \citealt{2014ApJ...781...41K}).
The parameters of the thermal component are the two plasma temperatures, $kT_s$ and $kT_h$, and the two ionization timescales, $\tau_s$ and $\tau_h$.

The fit with all the element abundances fixed at solar values \citep{1989GeCoA..53..197A} gives, for the three regions, best fits with $\chi^2/\mathrm{dof}=495.45/475$ ($6''-15''$), $1733.25/1254$ ($6''-48''$) and $2111.99/1414$ ($6''-60''$).
We then allowed the abundances of Mg, Si, and S --- which are responsible for the main features observed at energies $\approx\!1.25$, $\approx\!1.8$ and $\approx\!2.4\,\mathrm{keV}$, respectively --- to vary in the hotter plasma component.
This led to a significant improvement in the $6''-48''$ and $6''-60''$ regions (F-test probability $<10^{-2}$), while the fit of the $6''-15''$ region did not change significantly (F-test probability $=0.38$). 

These models (Figure~\ref{fig:snr_spec} and the best fit parameters in Table~\ref{tab:snr}), properly rescaled to account for the excision of the inner $6^{\prime\prime}$ circle, were then added as fixed components in the spectral fits of the data from \xmm, \ixpe\ and \nustar\ (\S \ref{subsec:pre} and \S \ref{subsec:post}).

\subsection{\xmm}
\label{subsec:xmm}

We analyzed the latest and deepest \xmm\ observation of \src\ (see Table~\ref{tab:log}). The analysis of this dataset is presented in this \emph{Letter} for the first time.
The European Photon Imaging Cameras (EPIC) instrument was operated in Prime Full Window (pn camera, \citealt{2001A&A...365L..18S}) and Prime Partial Window (MOS1/2 cameras, \citealt{2001A&A...365L..27T}) with the thick filter.

Data reduction was performed using the \texttt{epproc} and \texttt{emproc} pipelines of version 18 of the Science Analysis System (SAS).
The extraction regions for the source and background are shown in Figure \ref{fig:snr}.
We selected single- and multiple-pixel  events (\texttt{pattern}$\leq$4 for the EPIC-pn and $\leq$12 for the EPIC-MOS).

\subsection{\nustar}
\label{subsec:nustar}

\nustar\ \citep{2013ApJ...770..103H} observed \src\ twice, following the recent bursting activity: the first observation started on August 29 and lasted $50\,\mathrm{ks}$; the second started on September 28 and lasted $56\,\mathrm{ks}$. Furthermore, to compare the properties of broadband emission before and after the bursting episode, we analyzed a long observation ($100\,\mathrm{ks}$) made in September 2013 (see Table~\ref{tab:log}). Data were processed using the \texttt{nupipeline} tool with default screening parameters, and spectra were extracted throughout the energy bandpass using \texttt{nuproducts}. The extraction regions for the source and background are shown in Figure \ref{fig:snr}.

\section{Results} \label{sec:results}
We converted the times of arrival of the photons to the Solar System barycenter using the Jet Propulsion Laboratory Development Ephemeris JPL DE430, with source coordinates \radec.
Spectral and spectro-polarimetric analyses were performed using \texttt{XSPEC} \cite[][version 12.11.0]{Arnaud1996}. 
Absorption from the interstellar medium was accounted for using the \textsc{tbabs} models with cross sections and abundances from \citet{1989GeCoA..53..197A}.
The $\chi^2$ statistics was used to assess the goodness of the fits, and all reported errors are at $1\sigma$ confidence level (cl hereafter), unless otherwise specified.

\subsection{Spectral analysis} \label{subsec:spectral}

\subsubsection{Pre bursting activity} 
\label{subsec:pre}

We simultaneously fit the \xmm-EPIC and \nustar\ spectra of \src\ obtained before the August 2024 bursting phase.
The emission component of SNR Kes 73 was included as explained in \S \ref{subsec:cha}, and we added a constant to account for the calibration uncertainties between the three EPIC cameras and the two \nustar\ modules. We found that these normalization parameters for \nustar\ are 15\%--20\% higher than those of \xmm\ for all the tested models.

\begin{figure*}[ht]
\begin{center}
\includegraphics[trim={1cm 1cm 3cm 1cm},clip,height=7.5cm]{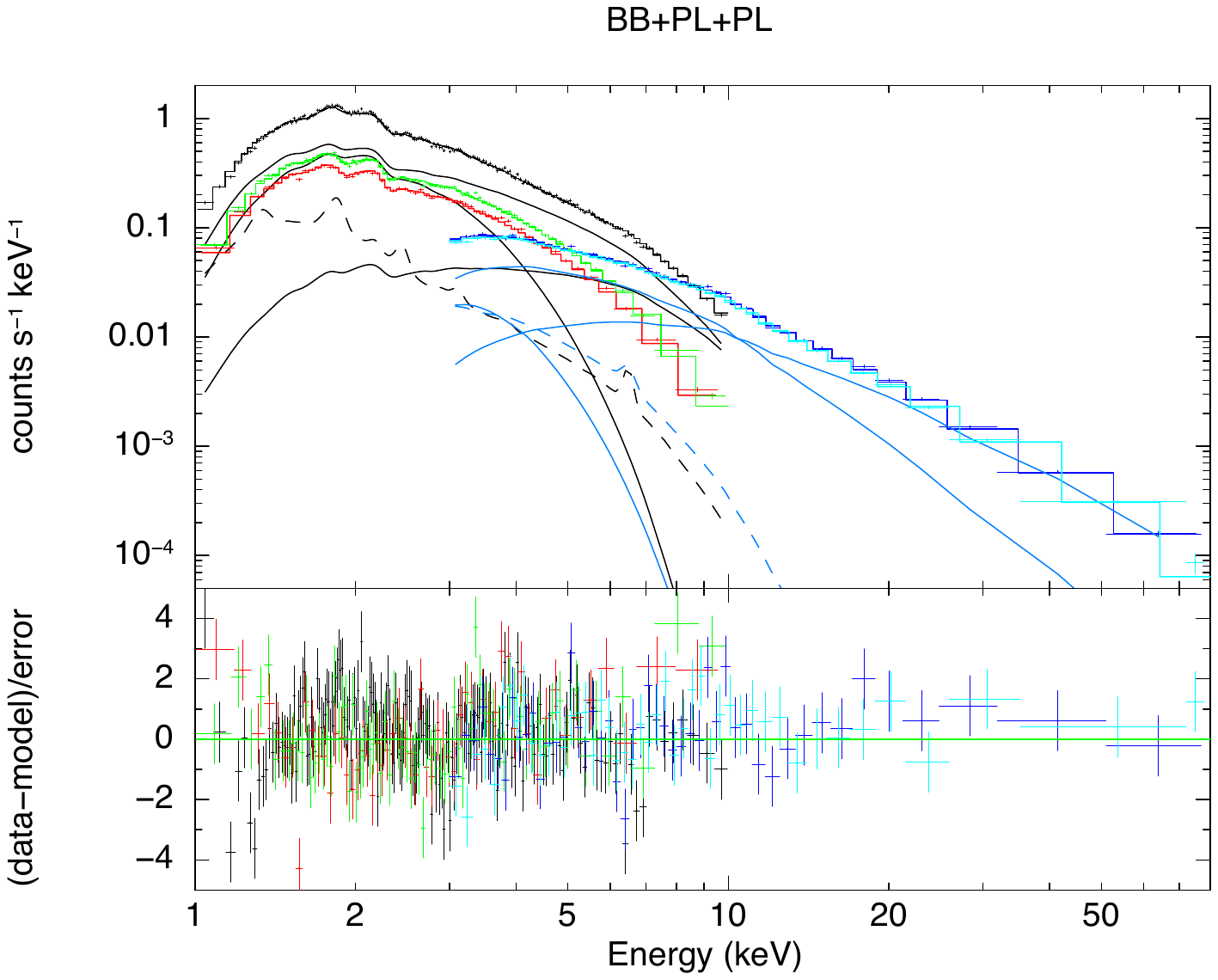}
\includegraphics[trim={2.5cm 1cm 3cm 1cm},clip,height=7.5cm]{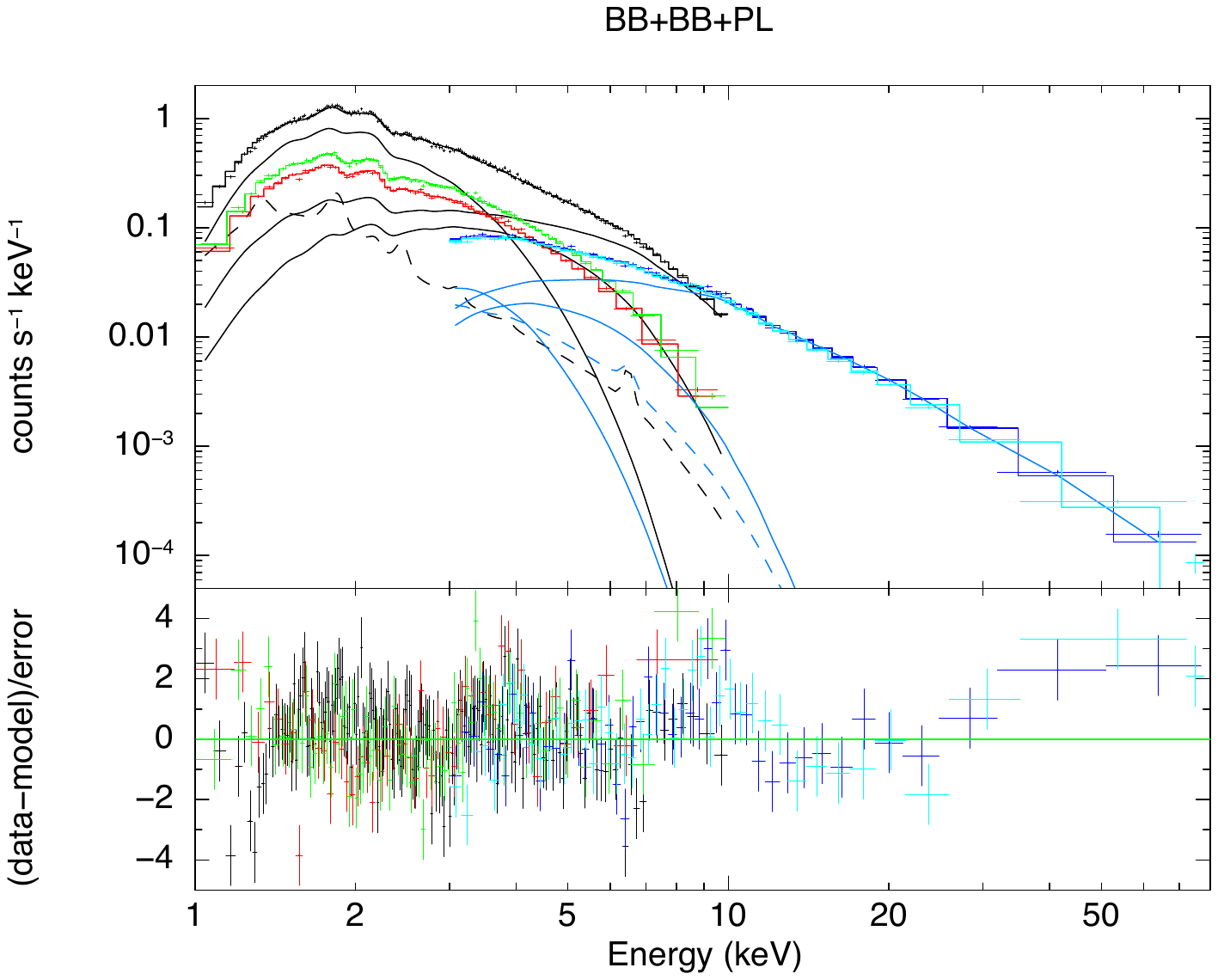}
\caption{Best fit and residuals to the \xmm\ (EPIC-pn in black, MOS1 and MOS2 in red and green respectively) and \nustar\ (FPMA and FPMB detectors in blue and cyan respectively) pre-burst data. 
The solid lines show the single components of the models, while the dashed lines show the SNR contribution as computed in \S\ref{subsec:cha}, which was kept fixed in all fits. The plotted spectra were rebinned to improve visualization and single components are only displayed in the cases of the EPIC-pn and FPMA detectors for clarity.
}
\label{fig:spec_pre}
\end{center}
\end{figure*}

We fitted to the spectra a (absorbed) two-component model, either two power laws (PL+PL), or a blackbody and a power law (BB+PL). In both cases, the fit was poor, with $\chi^2/\mathrm{dof} = 3319.90/2671$ and $3331.42/2671$, respectively, with a structured modulation in the residuals, suggesting the presence of an additional component. We then added a BB as a third component, obtaining $\chi^2/\mathrm{dof} = 2847.25/2669$ for the BB+PL+PL model and $\chi^2/\mathrm{dof} = 2919.53/2669$ for the BB+BB+PL model
(see the two panels in Figure~\ref{fig:spec_pre} and Table~\ref{tab:spec}). The addition of the third component yields in both cases an F-test probability $<$10$^{-16}$.
The total unabsorbed fluxes in the $2$--$8\,\mathrm{keV}$ energy range turned out to be $F_\mathrm{BB+PL+PL} = (2.11\pm0.01) \times 10^{-11}\,\flux$ and $F_\mathrm{BB+BB+PL} = (2.03\pm0.01) \times 10^{-11}\,\flux$.

\subsubsection{Post bursting activity} 
\label{subsec:post}

We fit the simultaneous \ixpe\ and \nustar\ spectra in the $2$--$79\,\mathrm{keV}$ energy range. A cross-calibration constant was introduced to account for the different response of the three \ixpe\ DUs and the two \nustar\ modules; we found that these normalization parameters for \nustar\ are $\approx\!30$\% higher than those of \ixpe\ for all the tested models. The contribution of SNR Kes 73 was included as explained in \S \ref{subsec:cha}. 

We started with absorbed PL+PL and BB+PL models and obtained best fits with $\chi^2/\mathrm{dof}=904.11/883$ and $972.07/883$, respectively. The PL+PL fit is formally acceptable, but yields a very high column absorption $\nh = (4.4\pm0.2)\times 10^{22}\,\mathrm{cm}^{-2}$ and an unreasonably steep photon index of $\Gamma_{\rm soft}= 5.00\pm0.14$, probably mimicking the presence of a thermal component.
If we fix the absorption to the value used by \citet{2024arXiv241216036S}, $\nh=2.6\times10^{22}$ cm$^{-2}$, we obtain $\Gamma_{\rm soft}= 3.72\pm0.04$ and $\Gamma_{\rm hard}= 1.21\pm0.02$, which are quite in agreement with what found by \citet{2024arXiv241216036S}. However, this fit is marginally acceptable ($\chi^2/\mathrm{dof} = 1007.90/884$) and the addition of the thermal component is statistically required (F-test probability $\approx2\times10^{-21}$, see below).

\begin{figure*}[ht]
\begin{center}
\includegraphics[trim={1cm 1cm 3cm 1cm},clip,height=7.5cm]{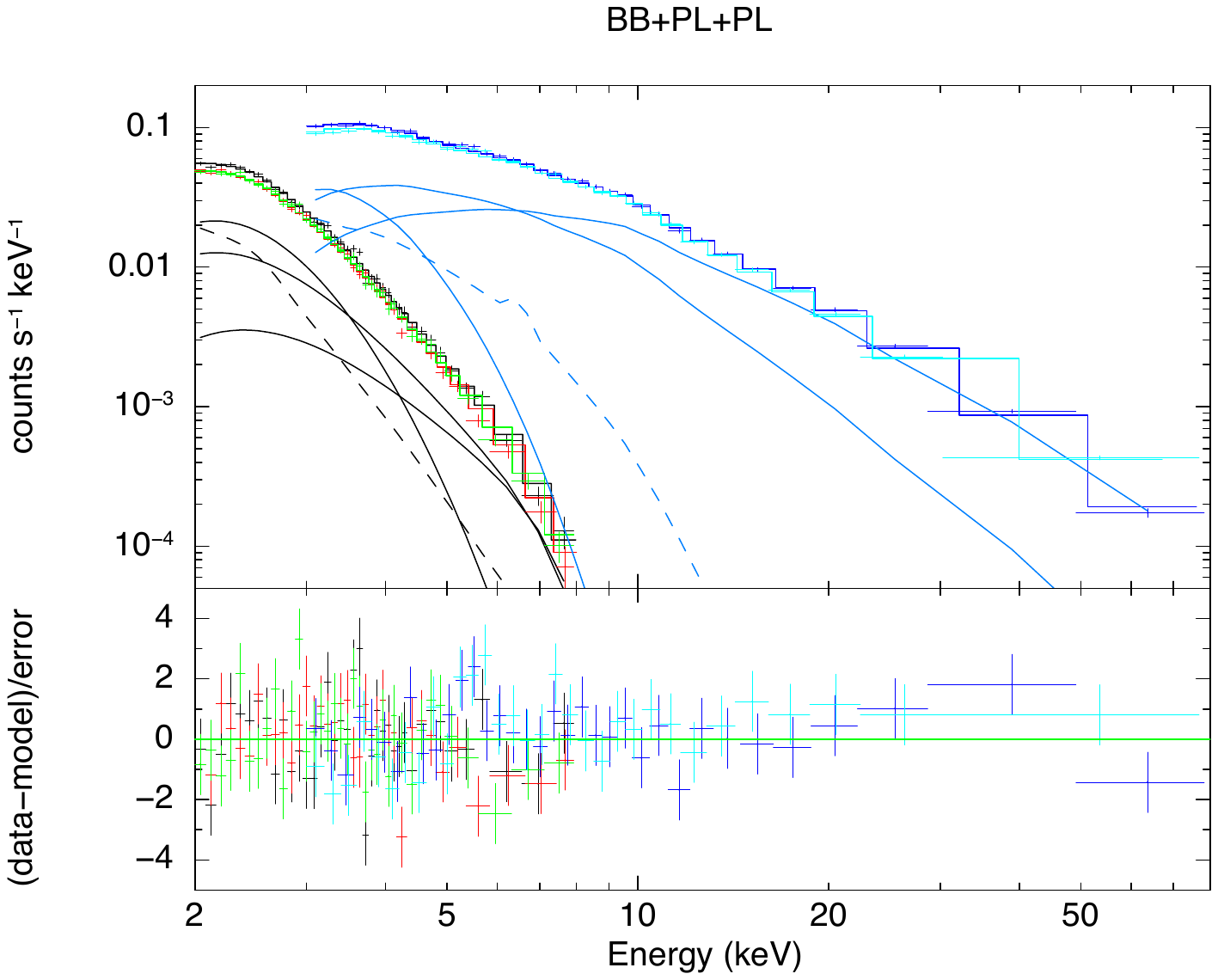}
\includegraphics[trim={2.5cm 1cm 3cm 1cm},clip,height=7.5cm]{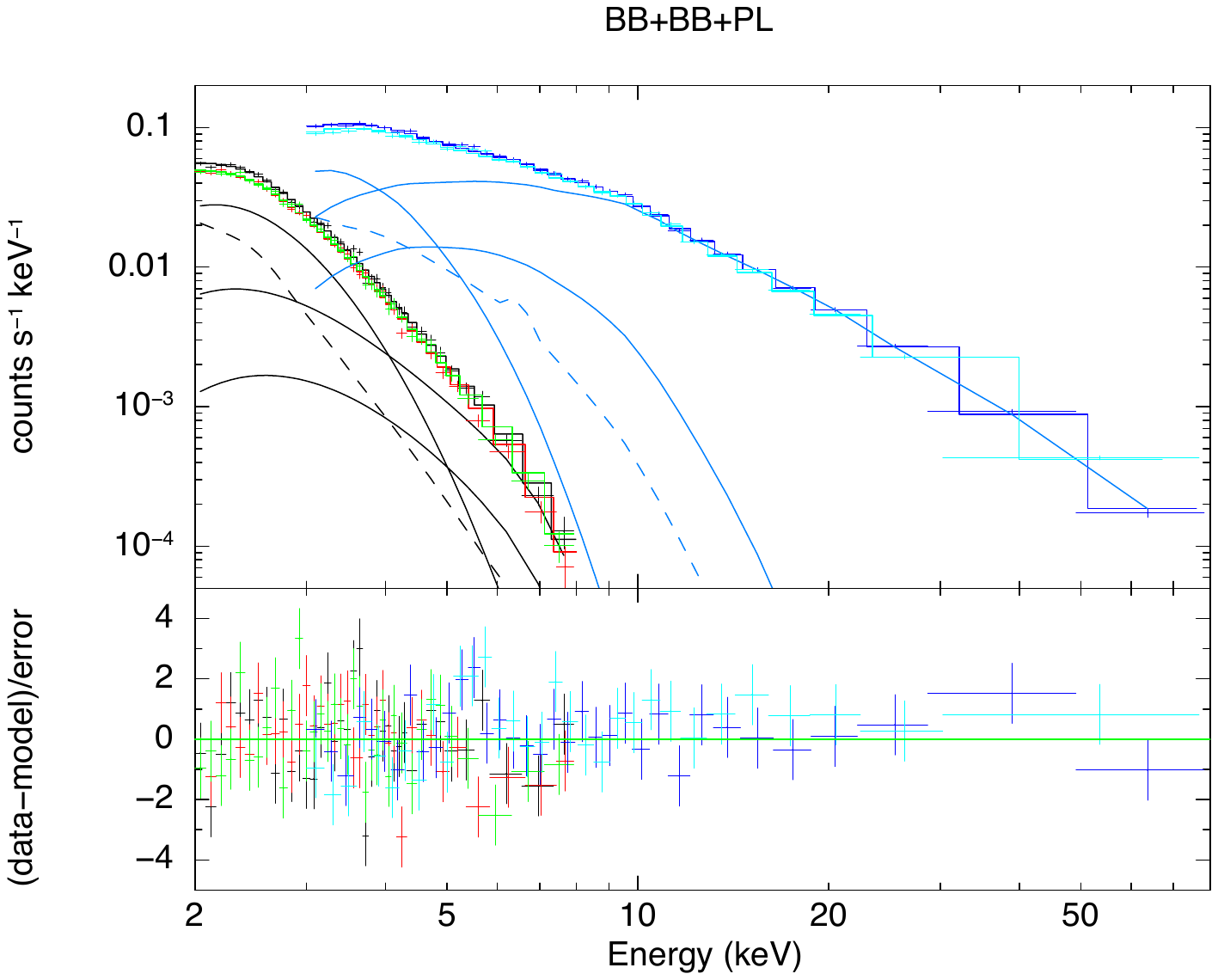}
\caption{Best fit and residuals to the \ixpe\ (DU1, DU2 and DU3 in black, red and green respectively) and \nustar\ (FPMA and FPMB detectors in blue and cyan respectively) 2024 data. Line code is as in Figure \ref{fig:spec_pre}.
\label{fig:spec_post}}
\end{center}
\end{figure*}

As a consequence, following the approach discussed in \ref{subsec:pre}, we considered a three-model spectral decomposition for the \ixpe\ and \nustar\ data. Given the reduced energy band of \ixpe\ and the lower number of source counts, we froze the value of the column density to those obtained in \S \ref{subsec:pre}.
We found that both the BB+PL+PL and BB+BB+PL decompositions provide an acceptable fit ($\chi^2/\mathrm{dof}=904.54/882$ and $900.77/882$, respectively)  with an even distribution of the residuals.
The data, together with the best-fit models, are shown in Figure \ref{fig:spec_post}, and the corresponding best-fit parameters are reported in Table \ref{tab:spec}. The unabsorbed fluxes in the $2$--$8\,\mathrm{keV}$ energy range are $F_\mathrm{BB+PL+PL} = (2.31\pm0.02) \times 10^{-11}\,\flux$ and $F_\mathrm{BB+BB+PL} = (2.22\pm0.02) \times 10^{-11}\,\flux$.

\begin{table*}[]
    \tabletypesize{\scriptsize}
    \begin{center}
    \caption{Fit parameters of \src\ spectra \label{tab:spec}}
    \begingroup
    \setlength{\tabcolsep}{10pt}
    \renewcommand{\arraystretch}{1.5}
    \begin{tabular}{l | c c | c c}
    \hline\hline
       & \multicolumn{2}{c}{BB+PL+PL} &  \multicolumn{2}{c}{BB+BB+PL}\\
    \hline
     & PRE & POST & PRE & POST \\
    \hline

   $\nh$ ($10^{22}\,\mathrm{cm}^{-2}$)  & $ 2.20_{-0.03}^{+0.03}$ & $2.20^*$ & $ 2.009_{-0.014}^{+0.015}$ & $2.009^*$ \\
   $kT_{\mathrm{BB}_1}$ (keV) & $ 0.491_{-0.004}^{+0.004}$ & $ 0.501_{-0.007}^{+0.007}$  & $ 0.465_{-0.008}^{+0.007}$ & $0.524_{-0.008}^{+0.008}$\\
   $R_{\mathrm{BB}_1}$ (km)  & $ 3.94_{-0.13}^{+0.12}$ & $ 4.62_{-0.23}^{+0.23}$ & $ 5.58_{-0.15}^{+0.18}$ & $ 4.65_{-0.15}^{+0.16}$ \\
   $F_\mathrm{BB_1}^{2-8}$ ($10^{-12}\,\flux$) & $ 5.25_{-0.25}^{+0.24}$ & $ 8.2_{-0.6}^{+0.5}$ & $ 7.6_{-0.4}^{+0.3}$ & $10.6_{-0.3}^{+0.3}$ \\

   $kT_{\mathrm{BB}_2}$ (keV) & -- & -- & $ 1.04_{-0.05}^{+0.05}$ &  $1.4_{-0.1}^{+0.1}$ \\
   $R_{\mathrm{BB}_2}$ (km) & -- & -- & $ 0.56_{-0.06}^{+0.07}$ & $ 0.24_{-0.03}^{+0.04}$ \\
   $F_\mathrm{BB_2}^{2-8}$ ($10^{-12}\,\flux$) & -- & -- & $ 4.3_{-0.2}^{+0.2}$ & $ 2.6_{-0.3}^{+0.3}$ \\

   $\Gamma_\mathrm{PL_2}$ & $ 2.5_{-0.1}^{+0.1}$ & $ 2.4_{-0.3}^{+0.2}$ & -- & -- \\
   $F_\mathrm{PL_2}^{2-8}$ ($10^{-12}\,\flux$) & $12.2_{-0.4}^{+0.4}$ & $ 9.3_{-0.9}^{+0.8}$ & -- & -- \\

   $\Gamma_\mathrm{PL_3}$ & $ 0.93_{-0.07}^{+0.06}$ & $ 1.1_{-0.07}^{+0.06}$ & $ 1.34_{-0.02}^{+0.02}$ & $ 1.30_{-0.03}^{+0.03}$\\
   $F_\mathrm{PL_3}^{2-8}$ ($10^{-12}\,\flux$) & $3.3_{-0.5}^{+0.6}$ & $ 5.7_{-1.7}^{+1.1}$ & $ 8.4_{-0.3}^{+0.3}$ & $ 9.0_{-0.4}^{+0.4}$ \\

   $F_\mathrm{TOT}^{2-8}$ ($10^{-12}\,\flux$) & $21.1_{-0.1}^{+0.1}$ & $23.1_{-0.2}^{+0.2}$ & $20.3_{-0.1}^{+0.1}$ & $22.2_{-0.2}^{+0.2}$ \\

   $F_\mathrm{TOT}^{8-79}$ ($10^{-12}\,\flux$) & $54.3_{-0.7}^{+0.7}$ & $59_{-1}^{+1}$ & $50.0_{-0.7}^{+0.7}$ & $58_{-1}^{+1}$ \\

   $K_{\rm pn}$ or $K_{\rm DU1}$ & $1^*$ & $1^*$ & $1^*$ & $1^*$ \\
   $K_{\rm MOS1}$ or $K_{\rm DU2}$ & $1.139\pm0.005$ & $0.972\pm0.011$ & $1.139\pm0.005$ & $0.971\pm0.011$ \\
   $K_{\rm MOS2}$ or $K_{\rm DU3}$ & $1.061\pm0.005$ & $0.978\pm0.011$ & $1.061\pm0.005$ & $0.978\pm0.011$ \\

   $K_{\rm FMPA}$ & $1.185\pm0.007$ & $1.328\pm0.017$ & $1.183\pm0.007$ & $1.326\pm0.017$ \\
   $K_{\rm FMPB}$ & $1.147\pm0.007$ & $1.314\pm0.017$ & $1.143\pm0.007$ & $1.312\pm0.017$ \\

   $\chi^2/\mathrm{dof}$ & 2847.25/2669  & 904.54/882 & 2919.53/2669 & 900.77/882\\

    \hline\hline
    \end{tabular}
    \endgroup
    \end{center}
    \tablecomments{
    The parameters marked with an asterisk were kept fixed in the fit. Blackbody radii are calculated assuming a distance of $8.5\,\mathrm{kpc}$ \cite[see][]{2008ApJ...677..292T}. The fluxes, corrected for the absorption, are in the specified energy range. $K$ are the cross-calibration constants between the \xmm-EPIC cameras and the \nustar\ FPM detectors (PRE), and between the \ixpe\ detector units and the \nustar\ FPM detectors (POST).
    }
\end{table*}

\subsection{Polarization analysis} \label{subsec:polarization}

\subsubsection{Energy-resolved polarimetric analysis} \label{subsubsec:polarizationenergy-resolved}
\begin{figure}[]
\begin{center}
\includegraphics[width=8.cm]{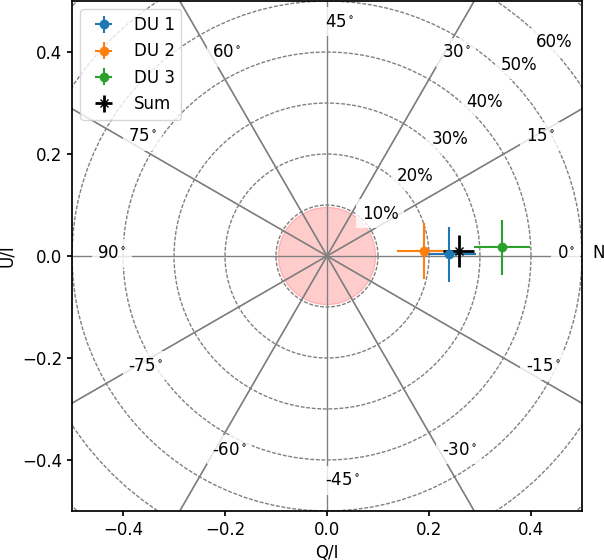}
\caption{The normalized Stokes parameters $Q/I$ and $U/I$ measured by the \ixpe\ DU1 (cyan), DU2 (orange) and DU3 (green) along with the combined measurements from the three DUs (black); $1\sigma$ error bars are also shown. Dashed circles represent the loci of constant $\mathrm{PD}$ while straight lines those of constant $\mathrm{PA}$. The $\mathrm{MDP}_{99}$ relative to the combined measurement is marked by the filled circle. \label{fig:qiuiplot}}    
\end{center}
\end{figure}

The phase-averaged Stokes parameters $I$, $Q$ and $U$ integrated over the $2$--$8\,\mathrm{keV}$ band were obtained from the photon lists for each of the \ixpe\ DUs using the \texttt{ixpeobssim} package
\cite[][]{Baldini2022}. The results for the normalized Stokes parameters $Q/I$ and $U/I$ are shown in Figure \ref{fig:qiuiplot}. The detection obtained from the sum of the three DUs is highly significant ($\gtrsim8\sigma$), yielding a polarization degree $\mathrm{PD}=\sqrt{Q^2+U^2}/I = 25.9\pm3.1\%$, well above the minimum detectable polarization at $99\%$ cl \cite[$\mathrm{MDP}_{99}$, see][]{Weisskopf2010} of $9.5\%$, and a  polarization angle $\mathrm{PA}=\arctan{(U/Q)}/2=1.1\pm3.5^\circ$, measured east of celestial north.

We then divided the $2$--$8\,\mathrm{keV}$ energy band into four energy bins and extracted the values of $\mathrm{PD}$ and $\mathrm{PA}$ in each of them. Polarization properties were derived using both the \texttt{pcube} algorithm within the \texttt{ixpeobssim} suite and by fitting simultaneously the $I$, $Q$ and $U$ spectra with \texttt{XSPEC} in each energy interval. In the latter case, 
we first fit the \ixpe\ spectrum (Stokes $I$) in the $2$--$8\,\mathrm{keV}$ band using a simple BB+PL model with $\nh$ frozen at the value listed in Table \ref{tab:spec} ($\chi^2/\mathrm{dof}=230.95/216$). Then,  we fit simultaneously the spectra of $I$, $U$, and $Q$ in each energy interval, keeping the parameters of BB and PL at their best-fit values found previously, and convolving the spectral model with a constant polarization (\textsc{polconst}).

\begin{figure}[]
\begin{center}
\hspace{-0.5cm}
\includegraphics[width=8.5cm]{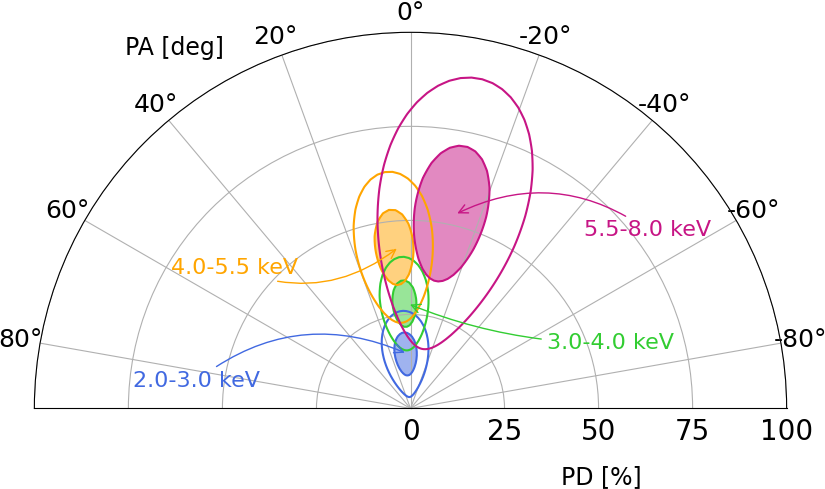}
\caption{Polar plot of the energy-dependent PD  and PA  obtained using \texttt{XSPEC} (see text for details). Filled and empty contours correspond to the $68\%$ and $99\%$ confidence regions, respectively, around the most probable values reported in Table \ref{tab:endeppol}. 
\label{fig:ovetti}  
}
\end{center}
\end{figure}

The results, reported in Table \ref{tab:endeppol}, show an overall agreement (within $1\sigma$) between the parameters obtained using the two methods, confirming a posteriori that the method used to extract $\mathrm{PD}$ and $\mathrm{PA}$ with \texttt{XSPEC} is essentially model independent. The measurements are above $\mathrm{MDP}_{99}$ in each energy bin, with significance $> 3\sigma$.
The most probable values of $\mathrm{PD}$ and $\mathrm{PA}$, together with the corresponding confidence contours at $68\%$ and $99\%$ cl, are also plotted in Figure \ref{fig:ovetti}. The degree of polarization increases monotonically with energy, from $\approx\!15\%$ at $2$--$3\,\mathrm{keV}$ up to more than $50\%$ at higher energies ($5.5$--$8\,\mathrm{keV}$), while the polarization angle remains fairly constant, with only a slight deviation in the bin at the highest energy. However, the polarization direction is compatible with that of the celestial north in the entire \ixpe\ band within $1\sigma$.
\begin{table*}[]
    \tabletypesize{\scriptsize}
    \begin{center}
    \caption{Energy-resolved polarization degree and angle\label{tab:endeppol}}
    \begingroup
    \setlength{\tabcolsep}{10pt}
    \renewcommand{\arraystretch}{1.5}
    \begin{tabular}{lccccc}
    \hline\hline
    \ &  $2.0$--$3.0\,\mathrm{keV}$ & $3.0$--$4.0\,\mathrm{keV}$ & $4.0$--$5.5\,\mathrm{keV}$ & $5.5$--$8.0\,\mathrm{keV}$ & $2.0$--$8.0\,\mathrm{keV}$\\
    \hline
    {\tt pcube} PD ($\%$) & $16.1\pm3.5$ & $28.0\pm4.0$ & $37.9\pm6.1$ & $54.2\pm15.2$ & $25.8\pm3.1$\\
    {\tt pcube} PA (deg) & $4.6\pm6.3$ & $2.1\pm4.1$ & $2.0\pm4.6$ & $-6.7\pm8.0$ & $1.1\pm3.5$\\
    {\tt pcube} PD S/N & $4.6\sigma$ & $7.0\sigma$ & $6.2\sigma$ & $3.6\sigma$ & $8.3\sigma$\\
    {\tt pcube} $\mathrm{MDP}_{99}$ ($\%$) & $10.7$ & $12.1$ & $18.5$ & $45.9$ & $9.5$\\
    \texttt{XSPEC} PD ($\%$) & $14.5\pm3.8$ & $27.9\pm4.1$ & $43.1\pm6.6$ & $52.9\pm12.1$ & $25.3\pm2.4$\\
    \texttt{XSPEC} PA (deg) & $3.3\pm7.5$ & $1.7\pm4.2$ & $4.1\pm4.4$ & $-12.9\pm6.7$ & $2.3\pm2.7$\\
    \texttt{XSPEC} PD S/N & $3.8\sigma$ & $6.8\sigma$ & $6.5\sigma$ & $4.4\sigma$ & $10.5\sigma$\\
    \hline\hline
    \end{tabular}
    \endgroup
    \end{center}
    \tablecomments{Values are extracted combining the contributions of the three \ixpe\ DUs, using the \texttt{ixpeobssim} and \texttt{XSPEC} suites (see text for details).
    Signal-to-noise (S/N) ratios are obtained dividing the most probable value by the corresponding $1\sigma$ error.}
\end{table*}

\subsubsection{Phase-resolved polarimetric analysis} \label{subsubsec:polarizationphase-resolved}
\begin{figure*}[]
\begin{center}
\includegraphics[width=18.cm]{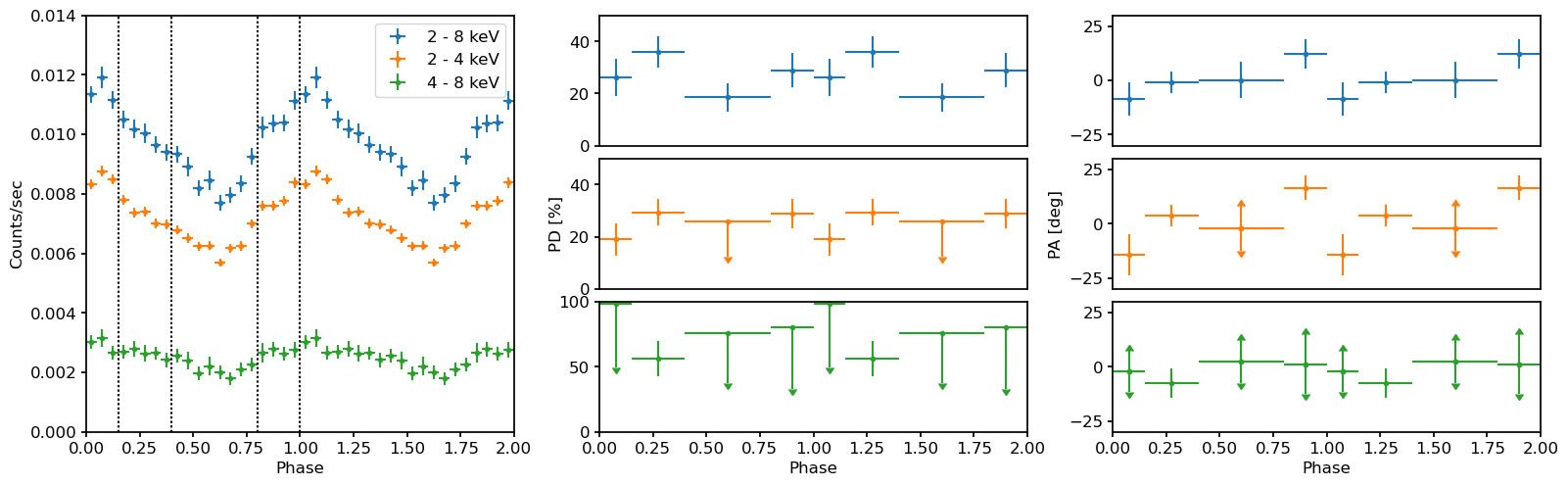}
\caption{\ixpe\ count rate (left panel), polarization degree (central panel) and polarization angle (right panel) as a function of the rotational phase; two cycles are shown for clarity. Results are presented for the $2$--$8$ (blue), $2$--$4$ (orange) and $4$--$8\,\mathrm{keV}$ (green) energy ranges. Polarization degree and angle are shown for the four phase intervals marked by vertical dotted lines in the left panel. In all the bins where PD is below $\mathrm{MDP}_{99}$, a $3\sigma$ upper limit (marked by a downward arrow) is shown for PD and the errors on PA are extended over the entire $[-90^\circ$,\,$90^\circ]$ range (double-headed arrows). \label{fig:phasedep}}   
\end{center}
\end{figure*}

To perform a phase-dependent polarimetric analysis, we measured the spin period of \src\ at the epoch of the \ixpe\ observation. Using only the \ixpe\ data, we determined the best period by searching around the frequency $0.085\,\mathrm{Hz}$ reported in \citet{DibKaspi2014} and using the $Z_n^2$-search technique included in the \texttt{HENDRICS} suite v.7.0 \cite[][]{bachettiHENDRICSHighENergy2018}. We found that the most likely value for the frequency (using $n=3$ harmonics, max $Z_3^2 = 850$) is $f=0.08469848(8)\,\mathrm{Hz}$, corresponding to a period $P=11.80659(1)\,\mathrm{s}$, at the MJD epoch $60587.140813793325$, while the value of the frequency derivative was not constrained.

The pulse profile shown in the left panel of Figure~\ref{fig:phasedep} is obtained folding the data at our best spin period and dividing the counts into $20$ equally spaced bins.
The pulse shape and the pulsed fraction\footnote{Defined as $\rm (max-min)/(max+min)$, where $\rm max$ and $\rm min$ are the maximum and the minimum count rates, respectively, of the pulse profile.} do not appear to change significantly with energy, with the latter being $(21.2\pm1.5)\%$ in the $2$--$4\,\mathrm{keV}$ band and only slightly higher (but still compatible within $1\sigma$) in the $4$--$8\,\mathrm{keV}$ range, $(27.6\pm7.3)\%$.

For our phase-dependent polarimetric analysis we consider four phase intervals, corresponding to the peak, decline, dip, and rise of the pulse (shown by the dotted vertical lines in the left panel of Figure~\ref{fig:phasedep}). This choice allows us to characterize the polarization properties in the relevant parts of the light curve and, at the same time, ensures a significant detection ($\mathrm{PD}>\mathrm{MDP}_{99}$) in each bin in the $2$--$8\ \mathrm{keV}$ range. However, the relatively short exposure time and the consequent choice of rather wide phase bins necessarily limits the inferences that can be drawn from such an analysis.
The results for $\mathrm{PD}$ and $\mathrm{PA}$, obtained using the \texttt{ixpeobssim}-\texttt{pcube} algorithm, are shown in Figure \ref{fig:phasedep} (center and right panels) and reported in Table \ref{tab:endeppol_Phase}.
In the $2$--$8\,\mathrm{keV}$ range, the polarization degree varies between $\approx\!20\%$ and $\approx\!36\%$. A fit with a simple sinusoidal profile of both counts and $\mathrm{PD}$ is acceptable around or better than $3\sigma$ cl and suggests that the two oscillations are in-phase within $1\sigma$ errors; fitting $\mathrm{PD}$ with a constant still produces an acceptable result, albeit only at $2\sigma$ cl. 

We also analyzed the phase-dependent behavior of the polarization in the $2$--$4$ and $4$--$8\,\mathrm{keV}$ energy ranges. At lower energies, the degree of polarization is above $\mathrm{MDP}_{99}$ in three phase bins and slightly below around the pulse minimum. If we assume the most probable value at the dip, $\mathrm{PD}$ still follows a sinusoidal pattern that is in phase with the light curve; a fit with a constant is ruled out at $96\%$ cl. At higher energies ($4$--$8\,\mathrm{keV}$), the polarization signal is significant in only one out of four phase intervals, excluding any conclusions.

Concerning the variation of $\mathrm{PA}$ with phase, both the fits with a constant and  a sinusoid in the $2$--$8\,\mathrm{keV}$ range are acceptable at better than $3\sigma$ cl, while those in the $2$--$4\,\mathrm{keV}$ range are both rejected. We then tried a fit in the $2$--$8\,\mathrm{keV}$ band using a rotating vector model \cite[RVM, see][]{rvm}. 
This is motivated by the fact that, if vacuum birefringence is indeed at work in magnetars, it is expected to produce phase-dependent variations in the polarization angle that match the RVM predictions -- a behavior already observed in a few sources \cite[see][]{taverna+22,zane+23,2024MNRAS.52712219H}. However, the two parameters of the model (the inclinations of the line-of-sight and magnetic axis with respect to the rotation axis) turned out to be totally unconstrained because of the limited statistics. In this respect, the analysis of the phase-dependent $\mathrm{PA}$ does not provide useful information on the source geometry, and the only conclusion that can be reached is that $\mathrm{PA}$ is not compatible with a constant in the soft energy band.

\setlength{\tabcolsep}{0.25em}
\begin{table}[]
    \tabletypesize{\scriptsize}
    \begin{center}
    \caption{Phase-resolved polarization degree and angle. \label{tab:endeppol_Phase}}
    \begin{tabular}{lcccc}
    \hline\hline
    Phase bin & Energy range &  PD  &  PA  &  PD S/N \\
    \ & (keV) & ($\%$) & (deg) & \\
    \hline
    & $2.0 - 8.0$ & $26.1 \pm 7.1$ & $-8.5\pm7.8$ & $3.7\sigma$ \\
    $0.0 - 0.15$ & $2.0 - 4.0$ & $19.0 \pm 6.3$ & $-14.3 \pm 9.4$ & $3.0\sigma$ \\ 
    & $4.0 - 8.0$ & $< 98.7$ & - & - \\ 
    \hline
    & $2.0 - 8.0$ & $35.8 \pm 6.0$ & $-0.9\pm4.8$ & $6.0\sigma$ \\
    $0.15 - 0.4$ & $2.0 - 4.0$ & $29.4 \pm 5.3$ & $3.7\pm5.1$ & $5.6\sigma$ \\ 
    & $4.0 - 8.0$ & $56.1 \pm 13.6$ & $-7.5\pm6.9$ & $4.1\sigma$ \\ 
    \hline
    & $2.0 - 8.0$ & $18.5 \pm 5.5$ & $0.2\pm8.4$ & $3.4\sigma$ \\
    $0.4 - 0.8$ & $2.0 - 4.0$ & $<25.7$ & - & - \\ 
    & $4.0 - 8.0$ & $<76.2$ & - & - \\ 
    \hline
    & $2.0 - 8.0$ & $29.0 \pm 6.8$ & $12.2\pm6.7$ & $4.3\sigma$ \\
    $0.8 - 1.0$ & $2.0 - 4.0$ & $28.9 \pm 5.7$ & $16.5\pm5.6$ & $5.1\sigma$ \\ 
    & $4.0 - 8.0$ & $<80.1$ & - & - \\
    \hline\hline
    
    \end{tabular}
    \end{center}
    \tablecomments{ 
    $3\sigma$ upper limits are provided when $\mathrm{PD}<\mathrm{MDP}_{99}$. S/N ratios are obtained dividing the most probable value by the corresponding $1\sigma$ error.}
\end{table}

\subsubsection{Spectro-polarimetric analysis} \label{subsubsec:specpolanalysis}
As a final step, we explored the polarization properties of the individual components entering the spectral models discussed in \S\ref{subsec:spectral},
i.e.\ {\sc tbabs} $\times$ $(${\sc snr} $+$ {\sc bbodyrad} $+$ {\sc bbodyrad} $+$ {\sc powerlaw}$)$ (hereafter model 1) and {\sc tbabs} $\times$ $(${\sc snr} $+$ {\sc bbodyrad} $+$ {\sc powerlaw} $+$ {\sc powerlaw}$)$ (hereafter model 2), where {\sc snr} stands for the SNR spectral model discussed in \S\ref{subsec:cha}. We froze all spectral parameters at the values reported in Table \ref{tab:spec} (POST) and convolved each spectral component with a constant polarization model ({\sc polconst}), simultaneously fitting the $I$, $Q$ and $U$ spectra.
\begin{figure}[]
\begin{center}
\includegraphics[width=9cm]{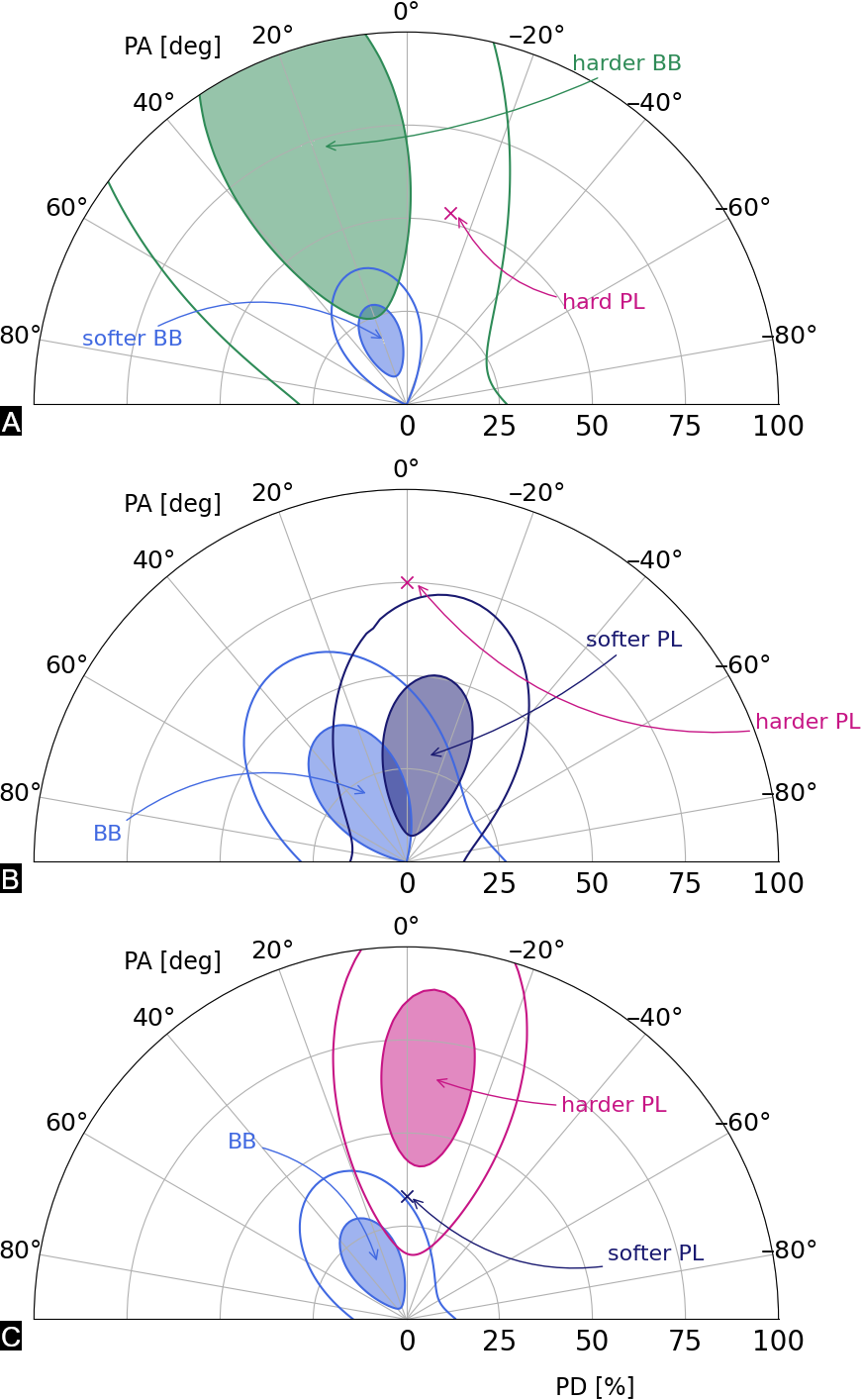}
\caption{Polarization of the single spectral components (see text for details).
Filled and empty contours correspond to the $68\%$ and $99\%$ confidence regions, respectively, around the most probable values; crosses mark the values at which $\mathrm{PD}$ and $\mathrm{PA}$ were frozen for the specified component. \label{fig:specpolmodels}}    
\end{center}
\end{figure}
Although the contribution of the SNR to the flux is sizable at $2$--$4\,\mathrm{keV}$ (see Figure \ref{fig:spec_post} and Table \ref{tab:snr}), its polarization is expected to be small, as discussed at the end of the section. 
Consequently, we froze the SNR polarization degree to zero. 

The hard PL accounts for almost all the counts above $\approx\!5\,\mathrm{keV}$ in the case of model 1 (see Figure~\ref{fig:spec_post}, right panel). Therefore, we fixed its polarization to the value measured with \ixpe\ in the $5.5$--$8\,\mathrm{keV}$ range (see Table \ref{tab:endeppol}) and allowed the polarization of the soft and hard BBs to vary. As shown in Figure \ref{fig:specpolmodels} (panel A), the polarization of the softer thermal component is approximately $20\%$ and is constrained at $99\%$ cl, while that of the hot blackbody is  unconstrained, 
and we can only conclude that the two thermal components are polarized in the same direction at a confidence level of $68\%$. 

On the other hand, in model 2 no single component dominates the spectrum in any energy range (see Figure~\ref{fig:spec_post}, left panel). 
So, none of them can be associated a priori with the polarization measured by \ixpe\ in a given energy interval.
Since a fit with all parameters left free is particularly poor, with all the parameters unconstrained, we decided to run different tests, allowing the fit to compute the polarization of two components and freezing that of the third one.

First, we checked that, by freezing only the degree of polarization of one of the three components and leaving all other polarization parameters free to vary, the polarization of the other two components is unconstrained at $68\%$. 
Then, for the component with frozen polarization, we fixed both $\mathrm{PD}$ and $\mathrm{PA}$. In particular, we decided to take $\mathrm{PA}=0^\circ$, in agreement, within $1\sigma$ cl, with the \ixpe\ measurement in all the energy bins (see Figure \ref{fig:ovetti}).

The high polarization degree obtained in the $2$--$8\,\mathrm{keV}$ energy interval suggests that, in the case of model 1, the hard PL component may be due to synchrotron radiation.
Hence, for model 2 we fixed the $\mathrm{PD}$ of the hard PL, taking $75\%$ as representative of synchrotron emission \cite[see e.g.][]{1979rpa..book.....R}. Note that this is not the PD measured in the $5.5$--$8\,\mathrm{keV}$ bin, since both PLs contribute in this energy range, allowing the soft PL to be less polarized. The result is reported in panel B of Figure \ref{fig:specpolmodels}. Although the polarization of both components turns out to be unconstrained at $99\%$ cl, the derived values at $68\%$ cl are $\approx\!20\%$ for the BB and $\approx\!30\%$ for the soft PL, while $\mathrm{PA}$ is compatible with zero, in agreement with observations.
The latter value of $\mathrm{PD}$ suggests that the soft PL may be produced by resonant Compton scattering \cite[RCS, which predicts $\mathrm{PD}\approx33\%$;][]{taverna+20}. We then froze this parameter at $33\%$, imposing again $\mathrm{PA}=0^\circ$, and leaving the polarization of the BB and the hard PL free to vary. The result is reported in Figure \ref{fig:specpolmodels} (panel C). In this case, the degree of polarization of the hard PL is well constrained at $\mathrm{PD}\approx65\%$ and its polarization angle is $\mathrm{PA}\approx-6^\circ$ (and compatible with $0^\circ$ within $1\sigma$). This suggests that, under the assumption that the polarization of each spectral component is constant across the \ixpe\ band, the data can be reproduced by a model comprising  a soft BB with $\mathrm{PD}\approx20\%$,  a mildly polarized soft PL ($\mathrm{PD}\approx30\%$) and a highly polarized hard PL, all with a polarization direction compatible with $0^\circ$.

Finally, we checked that by extracting the \ixpe\ counts in an annular region centered on the source, with inner and outer radii $30^{\prime\prime}$ and $150^{\prime\prime}$, respectively, from which the background was subtracted as described above, the {\tt pcube} PD turns out to be below $\mathrm{MDP}_{99}\approx20\%$, both in the $2$--$8\,\mathrm{keV}$ and in the $2$--$4\,\mathrm{keV}$ energy ranges. This supports the assumption that SNR emission exhibits relatively low polarization, consistent with other SNRs observed by \ixpe, where the average degree of polarization does not exceed $\approx\!10$\% at most \cite[see e.g.][]{2024Galax..12...59S}. As a further test, we performed the fit with model 2 under the same conditions discussed above, but considering the polarization angle frozen at $+90^\circ$ or $-90^\circ$. This resulted in a $>100\%$ polarization of at least one of the two free components, and thus reasonably rules out the possibility that the emission associated with a specific spectral component is polarized perpendicularly to that associated with the other two.

\section{Discussion} \label{sec:discuss}

In this work, we investigated the spectro-polarimetric properties of \src\ after it entered a period of bursting activity. We found that the emission over the whole \ixpe\ energy band is strongly polarized, with $\mathrm{PD}=(25.9\pm3.1)\%$ and $\mathrm{PA} =(1.1\pm3.5)^\circ$ at a significance greater than $8\sigma$. The degree of polarization increases monotonically with energy, and its variation with the rotational phase broadly follows the pulse profile. The polarization angle remains fairly constant in both energy and phase, with the exception of the $2$--$4\,\mathrm{keV}$ energy range, where a phase modulation in $\mathrm{PA}$ is detected.

The broadband ($2$--$79\,\mathrm{keV}$) spectrum, obtained joining \ixpe\ and simultaneous \nustar\ data, can be reproduced equally well by a BB+PL+PL or BB+BB+PL model. The $2$--$8\,\mathrm{keV}$ fluxes are about 10\% higher than those obtained with the same spectral decompositions from archival \xmm\ and \nustar\ data collected before the bursting phase. The increase in flux is more pronounced (about $15\%$) at higher energies, between $8$ and $79\,\mathrm{keV}$\footnote{
Due to the $\approx\!30$\% cross-correlation uncertainty between \ixpe\ and \nustar\ (see Table~\ref{tab:spec} and \citealt{2024arXiv241216036S}), the reported flux increments 
depend on our choice to consider \ixpe\ as the reference instrument; the different choice operated by \citet{2024arXiv241216036S} leads to a flux variation of 40\%.
}.

\begin{figure*}
\begin{center}
\includegraphics[width=0.24\textwidth]{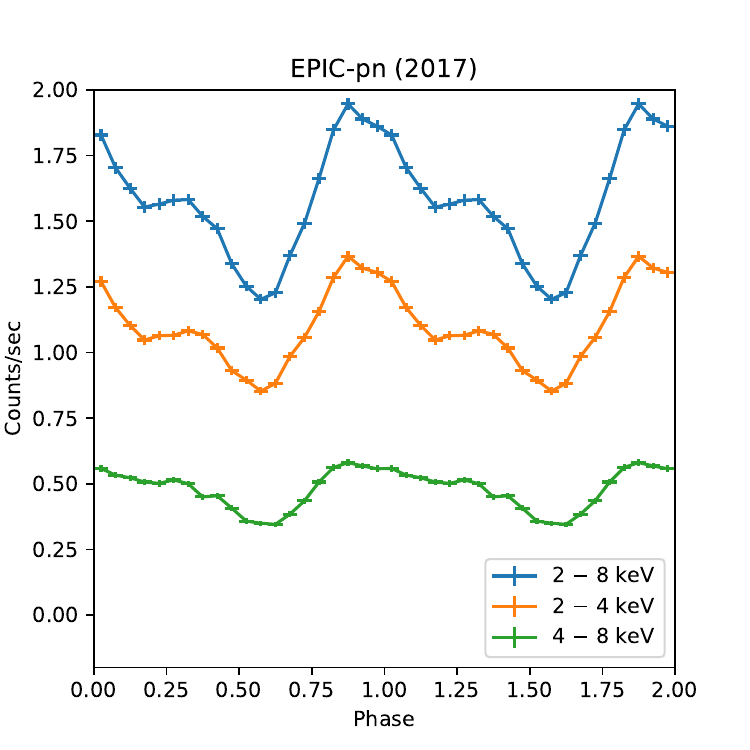}
\includegraphics[width=0.24\textwidth]{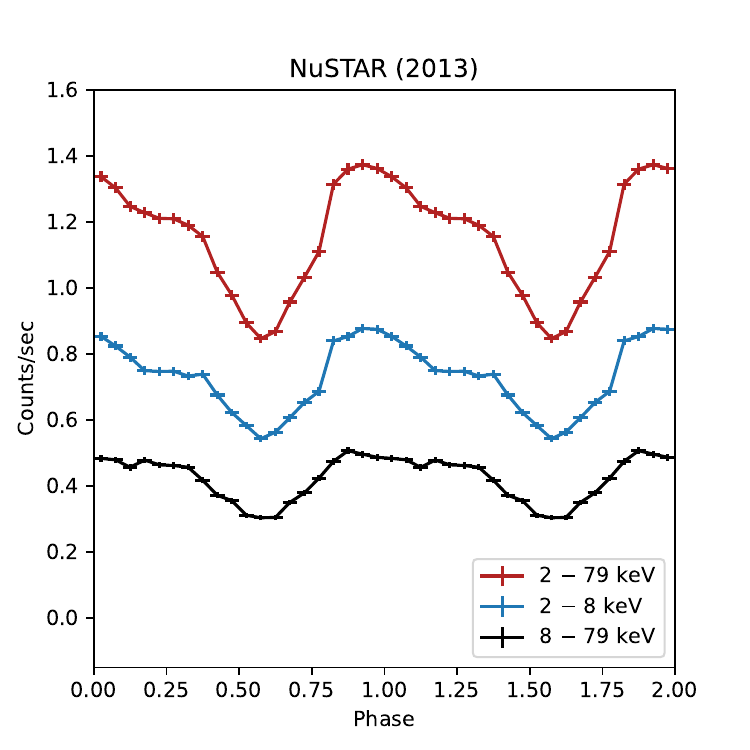}
\includegraphics[width=0.24\textwidth]{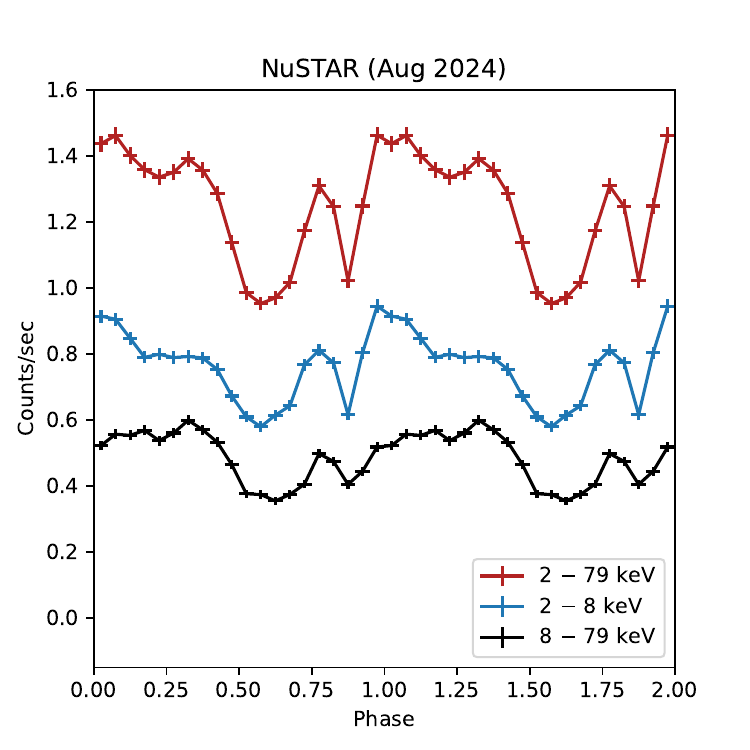}
\includegraphics[width=0.24\textwidth]{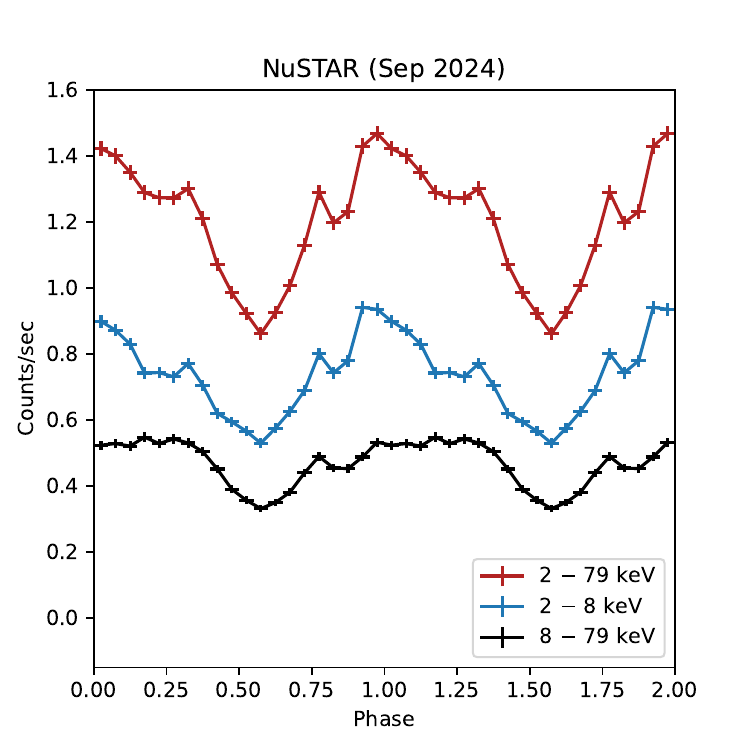}
\caption{Background-subtracted EPIC-pn pulse profiles in the $2$--$8$ (blue), $2$--$4$ (orange) and $4$--$8\,\mathrm{keV}$ (green) energy bands (left panel). The three rightmost panels show the \nustar\ pulse profiles in the $2$--$79$ (red), $2$--$8$ (blue) and $8$--$79\,\mathrm{keV}$ (black) energy bands at three different epochs.}
\label{fig:lc}
\end{center}
\end{figure*}
There is no substantial evolution of the spectral parameters with respect to the pre-burst state, although the temperature (radius) of the hotter BB increased (decreased) after the onset of the bursting activity. No evidence of a further hot and bright thermal component associated with a large heat deposition in the crust was found, as instead observed in full-fledged outbursts \cite[see e.g.][]{2018MNRAS.474..961C}.

However, a significant evolution is seen in the pulse profile. We found the best-folding spin period in each of the \xmm\ and \nustar\ observations (see Table~\ref{tab:log}) and we obtained the pulse profiles shown in Figure~\ref{fig:lc}.
Before the burst, the 2017 EPIC-pn and the 2013 \nustar\ data (first and second panels in Figure~\ref{fig:lc}) consistently show a double-peaked profile, with $\mathrm{PF} = (23.6\pm0.9)\%$ at $2$--$8\,\mathrm{keV}$ and $(25.2\pm1.6)\%$ at $8$--$79\,\mathrm{keV}$. In the \nustar\ data obtained a few days after the onset of the active burst phase, a dip appeared in the rise toward the primary peak (third panel, see also \citealt{2024ATel16802....1Y}) that partially recovered one month later (fourth panel).
The pulse profile measured by \ixpe\ (see Figure~\ref{fig:phasedep}, left panel) reveals the presence of two plateaux, just before and after the main peak, which are reminiscent of the secondary peak and the small dip seen in the \nustar\ 2024 data (fourth panel in Figure \ref{fig:lc}). These features are not well resolved in the \ixpe\ light curve, likely due to the lower number of events collected in the relatively short exposure time. The pulsed fraction attains a value quite similar to that of the other instruments, with $\mathrm{PF}=(21.5\pm2.2)\%$ in the $2$--$8\,\mathrm{keV}$ band, without significant changes with energy.

Like other magnetars, \src\ is known to exhibit significant hard X-ray emission. These high-energy tails were originally detected by INTEGRAL \citep{2004ApJ...613.1173K,2006ApJ...645..556K,2005A&A...433L...9M,2005A&A...433L..13M,2006A&A...449L..31G}, can extend up to a few hundreds of keV, and are highly phase dependent \citep{2008A&A...489..263D,2008A&A...489..245D,2014ApJ...789...75V,2016ApJ...831...80Y}. Their physical origin is still poorly understood and different interpretations have been proposed, ranging from thermal bremsstrahlung in the surface layers heated by returning currents, to synchrotron emission from pairs in the magnetosphere \citep{2005ApJ...634..565T}, to Compton resonant upscattering from relativistic charges \citep{2005ApJ...630..430B,2008AIPC..968...93B,2013ApJ...762...13B,2018ApJ...854...98W}. 
Thanks to \nustar, we were able to characterize this emission component after the beginning of the burst-active phase.
We find that the hard power law contributes significantly to the large polarization of the source at higher energies ($5$--$8\,\mathrm{keV}$). 

Attempts to perform a spectro-polarimetric analysis of the \ixpe\ data alone by associating a constant polarization model with each spectral component and leaving all parameters free to vary resulted in a largely unconstrained fit. However, fixing the polarization parameters of one of the spectral components produces a reasonably constrained fit, and the hard PL turns out to be polarized at more than $65\%$, regardless of which component is actually frozen. This large polarization is difficult to reconcile with resonant Compton scattering models, which predict $\mathrm{PD}\approx30\%$--$40\%$ because of the $1/3$ ratio of the X- and O-mode photon scattering cross sections \cite[][see also \citealt{2024arXiv241216036S}]{fernandez2011,taverna+14,2018ApJ...854...98W}. 
However, the polarization degree and the power-law spectral distribution may be compatible with a synchrotron/curvature origin for the high-energy tail.
In fact, if one assumes that synchrotron radiation is produced by a power-law distribution of electrons, the photon index $\Gamma_{\rm hard}=\Gamma_{\mathrm{PL}_3}\sim1$ (see Table \ref{tab:spec} POST) implies a degree of polarization of $\approx\!75\%$ \cite[][]{1979rpa..book.....R}.
Interestingly, by fixing the polarization degree of the hard PL at $75\%$, the soft PL turned out to be polarized at around $30\%$, as expected if it is produced by resonant Compton scattering in the star magnetosphere \cite[see][]{taverna+14,taverna+20}. Moreover, the polarization of the soft thermal component is $\approx\!25\%$, close to that observed in other magnetars \cite[][]{taverna+22,zane+23,2024MNRAS.52712219H}. Such polarization can hardly come from a strongly passive, magnetized atmosphere \cite[for $B\gtrsim 10^{14}\,\mathrm{G}$, also allowing for mode conversion;][]{2024MNRAS.528.3927K}, while it may be produced by a condensed surface or a bombarded atmosphere \cite[][]{taverna+20, kelly+24}.

Although the presence of a soft, thermal, mildly polarized component and a hard, strongly polarized power law appears to be well established, the nature of emission at intermediate energies ($\approx\!3$--$6\,\mathrm{keV}$) is not so clear. It could be either a softer power law, due to the RCS of the seed thermal photons, or a second, hotter thermal component. In the latter case, the relatively large polarization ($\approx\!30\%$) is incompatible with emission from a condensed surface and suggests that part of the star is covered by an atmosphere, as proposed for \srctwosh\ \citep[][]{zane+23}.

\begin{figure*}
\begin{center}
\includegraphics[width=1\textwidth]{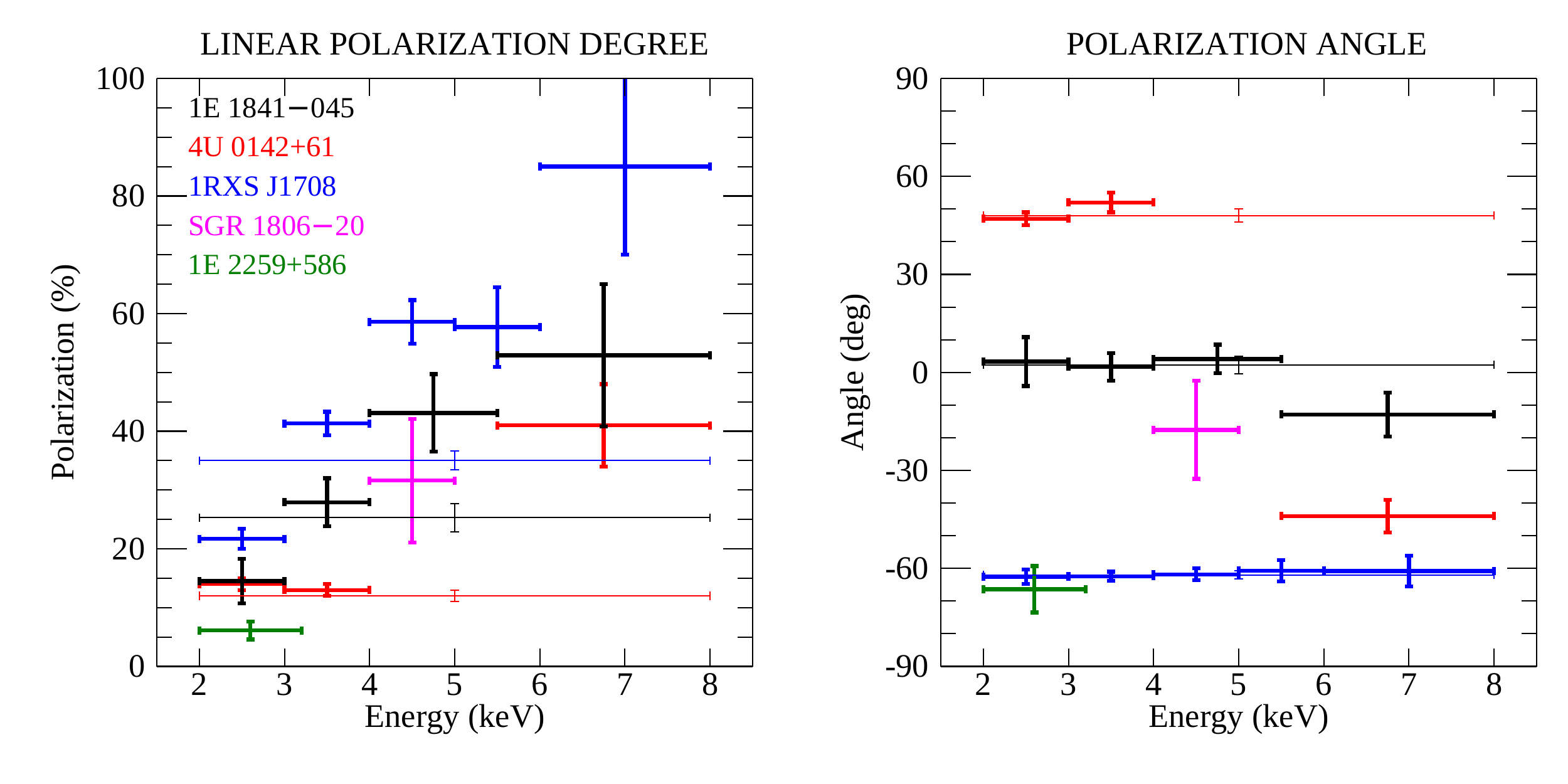}
\caption{Energy dependence of the linear polarization degree (left panel) and polarization angle (right panel) of the five magnetars observed with \ixpe. Thin lines indicate results for the total energy range (in case of significant polarization detection in the $2$--$8\,\mathrm{keV}$ range). See details in \citet{taverna+22,zane+23,2023ApJ...954...88T,2024MNRAS.52712219H}.}
\label{fig:magnetars}
\end{center}
\end{figure*}

Because of vacuum birefringence, the polarization angle in magnetars is actually fixed at the polarization limiting radius \cite[$r_\mathrm{pl}\approx80\,(B/10^{14}\, \mathrm{G})^{2/5}(E/1\, \mathrm{keV})^{1/5}\, R_\mathrm{NS}$;][]{heyl2002}, at least when radiation is produced inside $r_\mathrm{pl}$. As a consequence, radiation polarized in a given direction (wrt the local $B$-field) at emission is characterized by the same (phase-averaged) $\mathrm{PA}$ at infinity, which is related to the projection of the dipole axis on the plane of the sky \cite[][]{taverna+15}. This implies that the observed $\mathrm{PA}$ for the two normal modes must differ by $90^\circ$. In particular, if $\mathrm{PA}$ is counted from the projection of the star rotation axis in the plane of the sky, it is $\mathrm{PA}=0^\circ/90^\circ$ for O-/X-mode photons.

Since in \src\ the polarization direction changes little with energy,
the three spectral components should be polarized in the same mode if they come from below $r_\mathrm{pl}$. Although there is no way to tell which is the dominant polarization mode from the data, we note that both atmospheric emission (if the preferred model is BB+BB+PL) and RCS (in case of the BB+PL+PL decomposition) result in an intermediate component mostly polarized in the X-mode. This lends support to the soft BB being polarized in the X-mode, too. This is compatible with the emission of a magnetic condensate which can be polarized in the X- or O-mode, depending on the emission/viewing geometry; the former was invoked for the thermal component in \srctwosh. Intriguingly, the polarization direction of the hard PL photons seems to be the same.

Synchrotron emission is polarized perpendicular to the magnetic field projection in the plane of the sky and is not, in general, associated with the X- and O-modes. However, if synchrotron radiation comes from below $r_\mathrm{pl}$, the strong $B$-field forces also synchrotron photons to propagate in the two normal modes, likely in the X-mode for the case at hand. 
A possibility is that synchrotron emission is due to a population of relativistic pairs produced \cite[similarly to what suggested by][]{2013ApJ...762...13B} at the base of a localized twisted bundle, where the magnetic field direction is expected to change little. In this picture, the phase-dependent polarization angle should oscillate according to the rotating vector model, with the amplitude depending on the geometry of the source. Unfortunately, as discussed above, not enough \ixpe\ counts have been collected to reconstruct the behavior of $\mathrm{PA}$ above $4\,\mathrm{keV}$.

The general behavior of the polarization in \src\ (monotonic increase with energy of $\mathrm{PD}$ at a constant polarization angle) is reminiscent of that observed in \srctwosh. The similarity between the two sources goes deeper, since both exhibit a double power-law tail with comparable spectral indices \citep[$\Gamma_\mathrm{soft}\sim 2.5$--$2.7$ and $\Gamma_\mathrm{hard}\sim 1$; see][for the X-ray spectrum of \srctwosh]{2008A&A...489..263D}. \cite{zane+23} proposed two possible scenarios for \srctwosh: either two thermal components, one with low polarization and the other with high polarization, or one thermal component polarized at $\approx\!20\%$ and a power law, with $\Gamma\sim 3$, polarized at $\approx\!70\%$. According to the latter interpretation, the same scenario discussed here for \src\ may also apply to \srctwosh, although the lack of simultaneous high-energy observations prevented in their case the complete characterization of the spectro-polarimetric properties at the upper end of the \ixpe\ energy range.

\section{Conclusions}

We have reported the first X-ray polarimetric measurements of the magnetar \src\ obtained with the \ixpe\ satellite, together with a detailed analysis of data from \cha, \xmm\ and \nustar. These complementary data have been essential to properly take into account the non-negligible contribution of the Kes 73 SNR (20\% of the $2$--$8\,\mathrm{keV}$ flux in the \ixpe\ source extraction region) and to place the observed broadband flux and spectral properties of \src\ in the context of the previous history of this persistent magnetar. In fact, the \ixpe\ observation was obtained 40 days after the period of bursting activity and the increase in persistent flux that occurred in August 2024.

The polarization properties found for \src\ are compared to those of the other four magnetars observed with \ixpe\ in Figure~\ref{fig:magnetars}. Similarly to \srctwosh, the linear polarization increases with the energy from $\approx\!15\%$ to $\approx\!55\%$, without any change in the polarization angle. 

We found that a three-component spectral model (BB+PL+PL or BB+BB+PL) is required to fit the simultaneous \ixpe\ and \nustar\ data, resulting in a flux of $8\times10^{-11}\,\flux$ between $2$--$79\,\mathrm{keV}$. The analysis of the archival \xmm\ and \nustar\ data obtained before the source activation favors a BB+PL+PL model with flux of $7.5\times10^{-11}\,\flux$ ($2$--$79\,\mathrm{keV}$, see also \citealt{2013ApJ...779..163A}).
Although only a moderate increase of the flux was found between the pre- and the post-burst phases, the pulse profile experienced a noticeable evolution, especially above $8\,\mathrm{keV}$ where the hard power law dominates the flux. 

High-energy tails are pretty ubiquitous in the magnetar population, but their origin is still poorly understood. Our spectro-polarimetric analysis showed that in \src\  this component is polarized at more than $65\%$ and can be well interpreted in terms of synchrotron/curvature emission.
The intermediate power law component has a degree of polarization of $30\%$, consistent with the predictions for resonant Compton scattering in the magnetosphere, while the moderate polarization of the soft, thermal component ($\approx\!25\%$) may be produced by a condensed surface or a bombarded atmosphere.

Future polarimetric missions in hard X-rays \citep[like PHEMTO and ASTROMEV;][]{2021ExA....51.1143L, 2021ExA....51.1225D} will prove key in assessing the synchrotron/curvature origin of magnetar hard tails and shed light on the nature of the other 
spectral components.

\begin{acknowledgments}
We acknowledge financial support from INAF Fundamental Research Grants through the ``Magnetars'' Large Program (PI S.Mereghetti).
The work of RTa, RTu, LM, and FM is partially supported by the PRIN grant 2022LWPEXW of the Italian Ministry of University and Research (MUR). FCZ acknowledges support from a Ram\'on y Cajal fellowship (grant agreement RYC2021-030888-I). RK acknowledges support from the Science and Technology Facilities Council (STFC) for funding through a Ph.D. studentship. We thank the anonymous referee for her/his helpful comments.
\end{acknowledgments}

%

\facilities{\ixpe, \xmm\ (EPIC), \cha\ (ACIS), \nustar.}


\software{\texttt{Astropy} \citep{2013A&A...558A..33A,2018AJ....156..123A},  
         \texttt{ixpeobssim} \citep[\url{https://ixpeobssim.readthedocs.io/en},][]{Baldini2022}, 
         \texttt{SAS} (\url{https://www.cosmos.esa.int/web/xmm-newton/sas}),
          \texttt{XSPEC} \citep{Arnaud1996},
          \texttt{CIAO} \citep{2006SPIE.6270E..1VF}.
          }



\appendix

\section{SNR Kes 73}

\setcounter{table}{0}
\renewcommand\thetable{\Alph{section}\arabic{table}}
\setcounter{figure}{0}
\renewcommand\thefigure{\Alph{section}\arabic{figure}}

\begin{figure*}[h]
\begin{center}
\includegraphics[trim={1cm 1cm 3cm 1cm},clip,height=5cm]{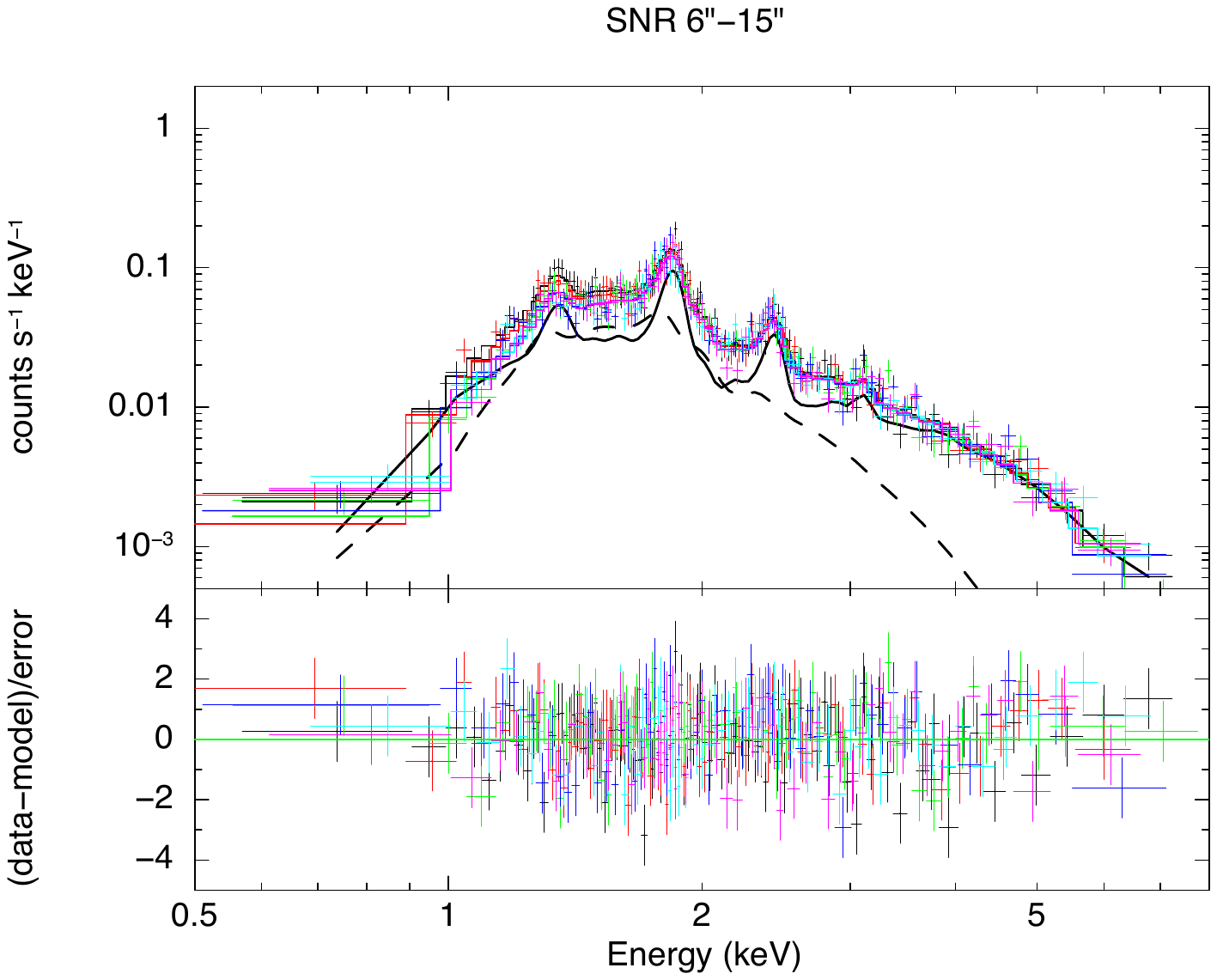}
\includegraphics[trim={2.5cm 1cm 3cm 1cm},clip,height=5cm]{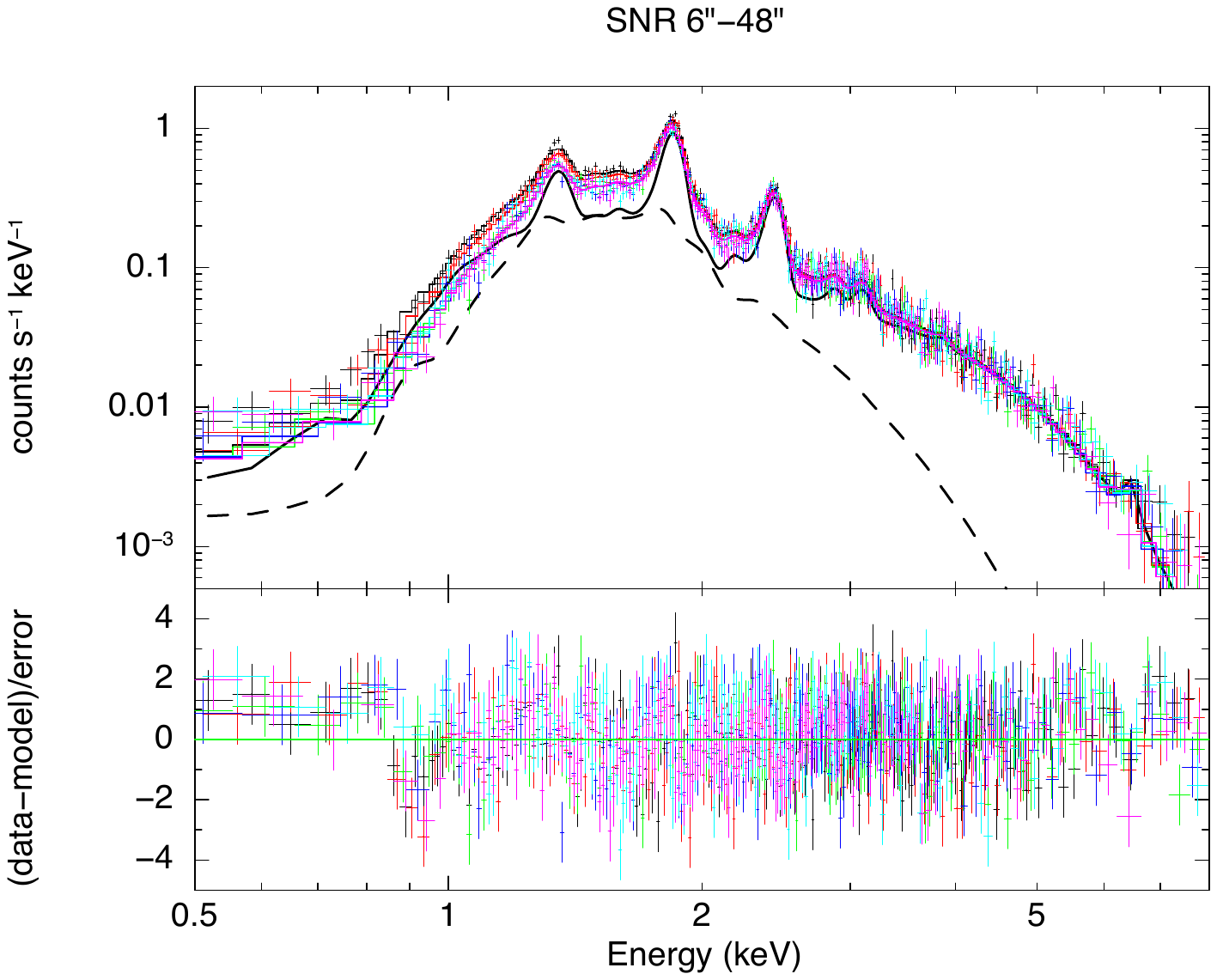}
\includegraphics[trim={2.5cm 1cm 3cm 1cm},clip,height=5cm]{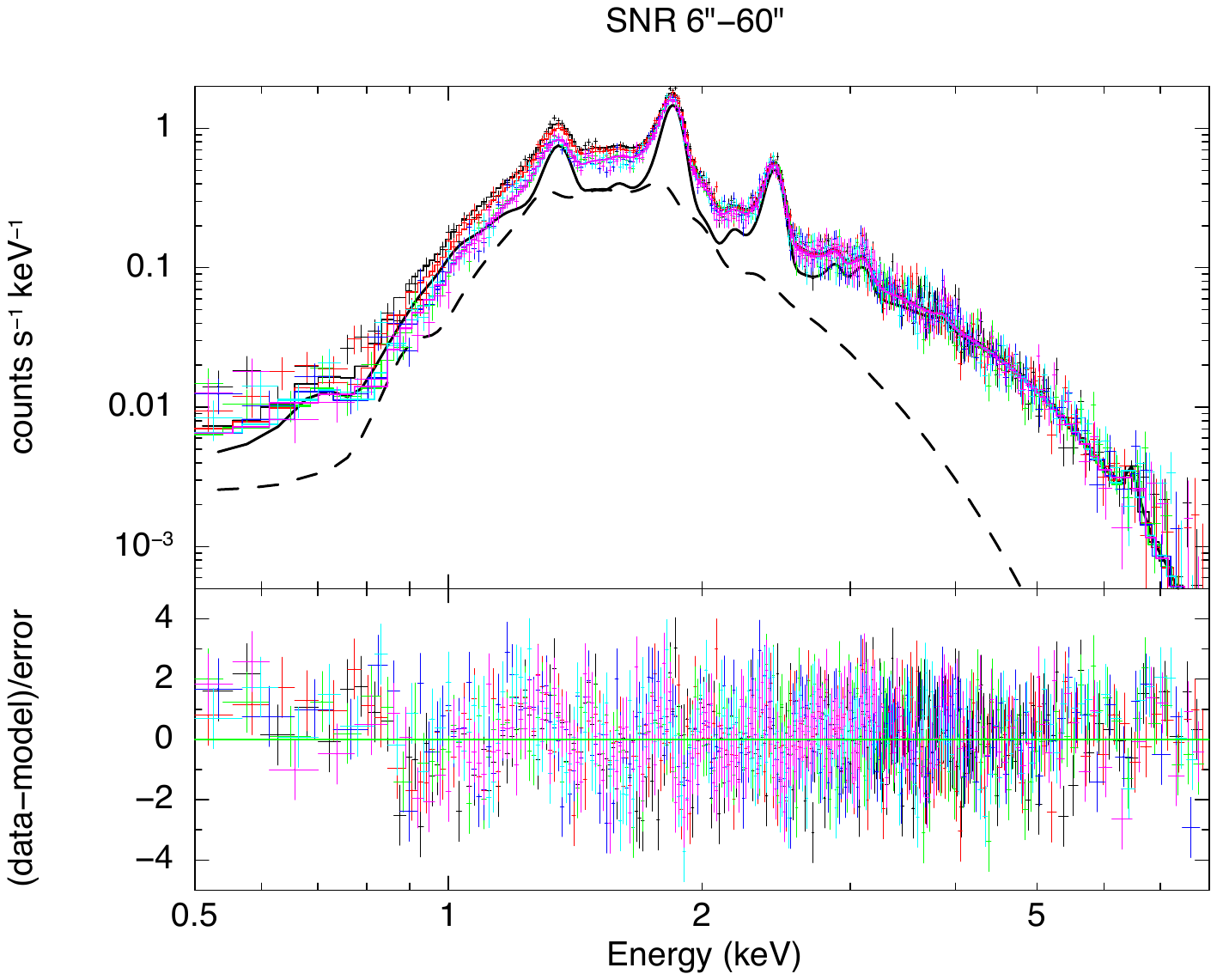}
\caption{Best fit and residuals to the \cha\ spectra of the SNR Kes 73. The three panels show the results for three annular extraction regions centered at (but excluding) \src, as detailed in \S\ref{subsec:cha}. The six observations, displayed in black, red, green, blue, cyan and magenta in increasing order of date, were jointly fit with two \textsc{vpshock} components, shown as a solid (hot component) and dashed (soft component) black lines.
\label{fig:snr_spec}}    
\end{center}
\end{figure*}

We report here the results of the spectral fit to the \cha\ data of SNR Kes 73 (see \S\ref{subsec:cha} for details). The best fit parameters are listed in Table \ref{tab:snr}; the spectrum together with the best-fit model is shown in Figure \ref{fig:snr_spec}.

\begin{table*}[h]
    \tabletypesize{\scriptsize}
    \begin{center}
    \caption{Fit parameters of SNR Kes 73 spectra \label{tab:snr}}
    \begingroup
    \setlength{\tabcolsep}{10pt}
    \renewcommand{\arraystretch}{1.5}
    \begin{tabular}{l c c c}
    \hline\hline
       & SNR $6''-15''$ & SNR $6''-48''$ & SNR $6''-60''$\\
     \hline

    $\nh$ ($10^{22}\,\mathrm{cm}^{-2}$)  & $2.31_{-0.05}^{+0.06}$ & $2.26_{-0.04}^{+0.06}$ & $2.32_{-0.03}^{+0.02}$ \\
    $kT_s$ (keV) & $0.61_{-0.04}^{+0.05}$ & $0.52_{-0.03}^{+0.05}$ & $0.49_{-0.01}^{+0.02}$\\
    $\tau_s$ ($10^{10}\,\mathrm{cm}^{-3}\,\mathrm{s}$)  & $0.15_{-0.08}^{+0.07}$ & $0.12_{-0.02}^{+0.02}$ & $0.14_{-0.01}^{+0.02}$\\
    $F_s^{2-8}$ ($10^{-12}\,\flux$) & $0.25_{-0.02}^{+0.02}$ & $1.21_{-0.06}^{+0.14}$ & $1.72_{-0.08}^{+0.06}$\\

    $kT_h$ (keV) & $3.2_{-0.2}^{+0.2}$ & $2.04_{-0.12}^{+0.04}$ & $1.79_{-0.02}^{+0.04}$\\
    $\tau_h$ ($10^{10}\,\mathrm{cm}^{-3}\,\mathrm{s}$)  & $3.9_{-0.4}^{+0.4}$ & $4.3_{-0.2}^{+0.3}$ & $4.9_{-0.2}^{+0.2}$\\
    $F_h^{2-8}$ ($10^{-12}\,\flux$) & $0.68_{-0.03}^{+0.03}$ & $3.67_{-0.12}^{+0.08}$ & $5.67_{-0.04}^{+0.18}$\\

    Mg  & $1^*$ & $1.38_{-0.10}^{+0.06}$ & $1.35_{-0.05}^{+0.05}$\\
    Si  & $1^*$ & $1.58_{-0.07}^{+0.05}$ & $1.61_{-0.05}^{+0.05}$\\
    S   & $1^*$ & $2.01_{-0.11}^{+0.08}$ & $2.05_{-0.08}^{+0.08}$\\

    $K$ & 1.19048 &      1.01587 &     1.01010  \\
    $F_{\rm TOT}^{2-8}$ ($10^{-12}\,\flux$) & $1.11_{-0.02}^{+0.02}$ & $4.98_{-0.03}^{+0.04}$ & $7.47_{-0.04}^{+0.05}$\\

    $\chi^2/\mathrm{dof}$ & 495.45/475  & 1537.00/1251 & 1813.83/1411 \\

    \hline\hline
    \end{tabular}
    \endgroup
    \end{center}
    \tablecomments{The parameters marked with an asterisk are kept fixed in the fit. All elemental abundances are in solar units as given by \citet{1989GeCoA..53..197A}. The fluxes, corrected for  absorption, are  in the $2$--$8\,\mathrm{keV}$ range; $F_s$ and $F_h$ are computed in the annular regions of radii $6''-15''$, $6''-48''$ and $6''-60''$, while $F_{\rm TOT}$ is corrected by the corresponding scaling factor $K$ to account for the  area lost in the inner circles. Errors are at $1\sigma$ cl level. }
\end{table*}

\bibliography{1841}{}

\begin{thebibliography}{}
\expandafter\ifx\csname natexlab\endcsname\relax\def\natexlab#1{#1}\fi
\providecommand{\url}[1]{\href{#1}{#1}}
\providecommand{\dodoi}[1]{doi:~\href{http://doi.org/#1}{\nolinkurl{#1}}}
\providecommand{\doeprint}[1]{\href{http://ascl.net/#1}{\nolinkurl{http://ascl.net/#1}}}
\providecommand{\doarXiv}[1]{\href{https://arxiv.org/abs/#1}{\nolinkurl{https://arxiv.org/abs/#1}}}

\bibitem[{{An} {et~al.}(2013){An}, {Hasco{\"e}t}, {Kaspi}, {Beloborodov},
  {Dufour}, {Gotthelf}, {Archibald}, {Bachetti}, {Boggs}, {Christensen},
  {Craig}, {Greffenstette}, {Hailey}, {Harrison}, {Kitaguchi}, {Kouveliotou},
  {Madsen}, {Markwardt}, {Stern}, {Vogel}, \& {Zhang}}]{2013ApJ...779..163A}
{An}, H., {Hasco{\"e}t}, R., {Kaspi}, V.~M., {et~al.} 2013, \apj, 779, 163,
  \dodoi{10.1088/0004-637X/779/2/163}

\bibitem[{{Anders} \& {Grevesse}(1989)}]{1989GeCoA..53..197A}
{Anders}, E., \& {Grevesse}, N. 1989, \gca, 53, 197,
  \dodoi{10.1016/0016-7037(89)90286-X}

\bibitem[{{Arnaud}(1996)}]{Arnaud1996}
{Arnaud}, K.~A. 1996, in Astronomical Society of the Pacific Conference Series,
  Vol. 101, Astronomical Data Analysis Software and Systems V, ed. G.~H.
  {Jacoby} \& J.~{Barnes}, 17

\bibitem[{{Astropy Collaboration} {et~al.}(2013){Astropy Collaboration},
  {Robitaille}, {Tollerud}, {Greenfield}, {Droettboom}, {Bray}, {Aldcroft},
  {Davis}, {Ginsburg}, {Price-Whelan}, {Kerzendorf}, {Conley}, {Crighton},
  {Barbary}, {Muna}, {Ferguson}, {Grollier}, {Parikh}, {Nair}, {Unther},
  {Deil}, {Woillez}, {Conseil}, {Kramer}, {Turner}, {Singer}, {Fox}, {Weaver},
  {Zabalza}, {Edwards}, {Azalee Bostroem}, {Burke}, {Casey}, {Crawford},
  {Dencheva}, {Ely}, {Jenness}, {Labrie}, {Lim}, {Pierfederici}, {Pontzen},
  {Ptak}, {Refsdal}, {Servillat}, \& {Streicher}}]{2013A&A...558A..33A}
{Astropy Collaboration}, {Robitaille}, T.~P., {Tollerud}, E.~J., {et~al.} 2013,
  \aap, 558, A33, \dodoi{10.1051/0004-6361/201322068}

\bibitem[{{Astropy Collaboration} {et~al.}(2018){Astropy Collaboration},
  {Price-Whelan}, {Sip{\H{o}}cz}, {G{\"u}nther}, {Lim}, {Crawford}, {Conseil},
  {Shupe}, {Craig}, {Dencheva}, {Ginsburg}, {VanderPlas}, {Bradley},
  {P{\'e}rez-Su{\'a}rez}, {de Val-Borro}, {Aldcroft}, {Cruz}, {Robitaille},
  {Tollerud}, {Ardelean}, {Babej}, {Bach}, {Bachetti}, {Bakanov}, {Bamford},
  {Barentsen}, {Barmby}, {Baumbach}, {Berry}, {Biscani}, {Boquien}, {Bostroem},
  {Bouma}, {Brammer}, {Bray}, {Breytenbach}, {Buddelmeijer}, {Burke},
  {Calderone}, {Cano Rodr{\'\i}guez}, {Cara}, {Cardoso}, {Cheedella}, {Copin},
  {Corrales}, {Crichton}, {D'Avella}, {Deil}, {Depagne}, {Dietrich}, {Donath},
  {Droettboom}, {Earl}, {Erben}, {Fabbro}, {Ferreira}, {Finethy}, {Fox},
  {Garrison}, {Gibbons}, {Goldstein}, {Gommers}, {Greco}, {Greenfield},
  {Groener}, {Grollier}, {Hagen}, {Hirst}, {Homeier}, {Horton}, {Hosseinzadeh},
  {Hu}, {Hunkeler}, {Ivezi{\'c}}, {Jain}, {Jenness}, {Kanarek}, {Kendrew},
  {Kern}, {Kerzendorf}, {Khvalko}, {King}, {Kirkby}, {Kulkarni}, {Kumar},
  {Lee}, {Lenz}, {Littlefair}, {Ma}, {Macleod}, {Mastropietro}, {McCully},
  {Montagnac}, {Morris}, {Mueller}, {Mumford}, {Muna}, {Murphy}, {Nelson},
  {Nguyen}, {Ninan}, {N{\"o}the}, {Ogaz}, {Oh}, {Parejko}, {Parley}, {Pascual},
  {Patil}, {Patil}, {Plunkett}, {Prochaska}, {Rastogi}, {Reddy Janga},
  {Sabater}, {Sakurikar}, {Seifert}, {Sherbert}, {Sherwood-Taylor}, {Shih},
  {Sick}, {Silbiger}, {Singanamalla}, {Singer}, {Sladen}, {Sooley},
  {Sornarajah}, {Streicher}, {Teuben}, {Thomas}, {Tremblay}, {Turner},
  {Terr{\'o}n}, {van Kerkwijk}, {de la Vega}, {Watkins}, {Weaver}, {Whitmore},
  {Woillez}, {Zabalza}, \& {Astropy Contributors}}]{2018AJ....156..123A}
{Astropy Collaboration}, {Price-Whelan}, A.~M., {Sip{\H{o}}cz}, B.~M., {et~al.}
  2018, \aj, 156, 123, \dodoi{10.3847/1538-3881/aabc4f}

\bibitem[{Bachetti(2018)}]{bachettiHENDRICSHighENergy2018}
Bachetti, M. 2018, Astrophysics Source Code Library, ascl:1805.019

\bibitem[{{Baldini} {et~al.}(2022){Baldini}, {Bucciantini}, {Di Lalla},
  {Ehlert}, {Manfreda}, {Omodei}, {Pesce-Rollins}, \& {Sgr{\`o}}}]{Baldini2022}
{Baldini}, L., {Bucciantini}, N., {Di Lalla}, N., {et~al.} 2022, SoftwareX, 19,
  101194, \dodoi{10.1016/j.softx.2022.101194}

\bibitem[{{Baring} {et~al.}(2005){Baring}, {Gonthier}, \&
  {Harding}}]{2005ApJ...630..430B}
{Baring}, M.~G., {Gonthier}, P.~L., \& {Harding}, A.~K. 2005, \apj, 630, 430,
  \dodoi{10.1086/431895}

\bibitem[{{Baring} \& {Harding}(2008)}]{2008AIPC..968...93B}
{Baring}, M.~G., \& {Harding}, A.~K. 2008, in American Institute of Physics
  Conference Series, Vol. 968, Astrophysics of Compact Objects, ed. Y.-F.
  {Yuan}, X.-D. {Li}, \& D.~{Lai} (AIP), 93--100, \dodoi{10.1063/1.2840459}

\bibitem[{{Barthelmy} {et~al.}(2015){Barthelmy}, {Kennea}, {Marshall},
  {Maselli}, \& {Sbarufatti}}]{2015GCN.18024....1B}
{Barthelmy}, S.~D., {Kennea}, J.~A., {Marshall}, F.~E., {Maselli}, A., \&
  {Sbarufatti}, B. 2015, GRB Coordinates Network, 18024, 1

\bibitem[{{Beloborodov}(2013)}]{2013ApJ...762...13B}
{Beloborodov}, A.~M. 2013, \apj, 762, 13, \dodoi{10.1088/0004-637X/762/1/13}

\bibitem[{{Caiazzo} {et~al.}(2022){Caiazzo}, {Gonz{\'a}lez-Caniulef}, {Heyl},
  \& {Fern{\'a}ndez}}]{Caiazzo2022}
{Caiazzo}, I., {Gonz{\'a}lez-Caniulef}, D., {Heyl}, J., \& {Fern{\'a}ndez}, R.
  2022, \mnras, 514, 5024, \dodoi{10.1093/mnras/stac1571}

\bibitem[{{Coti Zelati} {et~al.}(2018){Coti Zelati}, {Rea}, {Pons}, {Campana},
  \& {Esposito}}]{2018MNRAS.474..961C}
{Coti Zelati}, F., {Rea}, N., {Pons}, J.~A., {Campana}, S., \& {Esposito}, P.
  2018, \mnras, 474, 961, \dodoi{10.1093/mnras/stx2679}

\bibitem[{{De Angelis} {et~al.}(2021){De Angelis}, {Tatischeff}, {Argan},
  {Brandt}, {Bulgarelli}, {Bykov}, {Costantini}, {Curado da Silva}, {Grenier},
  {Hanlon}, {Hartmann}, {Hernanz}, {Kanbach}, {Kuvvetli}, {Laurent},
  {Mazziotta}, {McEnery}, {Morselli}, {Nakazawa}, {Oberlack}, {Pearce}, {Rico},
  {Tavani}, {Ballmoos}, {Walter}, {Wu}, {Zane}, {Zdziarski}, \&
  {Zoglauer}}]{2021ExA....51.1225D}
{De Angelis}, A., {Tatischeff}, V., {Argan}, A., {et~al.} 2021, Experimental
  Astronomy, 51, 1225, \dodoi{10.1007/s10686-021-09706-y}

\bibitem[{{den Hartog} {et~al.}(2008{\natexlab{a}}){den Hartog}, {Kuiper}, \&
  {Hermsen}}]{2008A&A...489..263D}
{den Hartog}, P.~R., {Kuiper}, L., \& {Hermsen}, W. 2008{\natexlab{a}}, \aap,
  489, 263, \dodoi{10.1051/0004-6361:200809772}

\bibitem[{{den Hartog} {et~al.}(2008{\natexlab{b}}){den Hartog}, {Kuiper},
  {Hermsen}, {Kaspi}, {Dib}, {Kn{\"o}dlseder}, \&
  {Gavriil}}]{2008A&A...489..245D}
{den Hartog}, P.~R., {Kuiper}, L., {Hermsen}, W., {et~al.} 2008{\natexlab{b}},
  \aap, 489, 245, \dodoi{10.1051/0004-6361:200809390}

\bibitem[{{Di Marco} {et~al.}(2023){Di Marco}, {Soffitta}, {Costa},
  {Ferrazzoli}, {La Monaca}, {Rankin}, {Ratheesh}, {Xie}, {Baldini}, {Del
  Monte}, {Ehlert}, {Fabiani}, {Kim}, {Muleri}, {O'Dell}, {Ramsey}, {Rubini},
  {Sgr{\`o}}, {Silvestri}, {Tennant}, \& {Weisskopf}}]{2023AJ....165..143D}
{Di Marco}, A., {Soffitta}, P., {Costa}, E., {et~al.} 2023, \aj, 165, 143,
  \dodoi{10.3847/1538-3881/acba0f}

\bibitem[{{Dib} \& {Kaspi}(2014)}]{DibKaspi2014}
{Dib}, R., \& {Kaspi}, V.~M. 2014, \apj, 784, 37,
  \dodoi{10.1088/0004-637X/784/1/37}

\bibitem[{{Duncan} \& {Thompson}(1992)}]{td92}
{Duncan}, R.~C., \& {Thompson}, C. 1992, \apjl, 392, L9, \dodoi{10.1086/186413}

\bibitem[{{Esposito} {et~al.}(2021){Esposito}, {Rea}, \&
  {Israel}}]{2021ASSL..461...97E}
{Esposito}, P., {Rea}, N., \& {Israel}, G.~L. 2021, in Astrophysics and Space
  Science Library, Vol. 461, Timing Neutron Stars: Pulsations, Oscillations and
  Explosions, ed. T.~M. {Belloni}, M.~{M{\'e}ndez}, \& C.~{Zhang}, 97--142,
  \dodoi{10.1007/978-3-662-62110-3_3}

\bibitem[{{Fermi GBM Team}(2024)}]{2024GCN.37234....1F}
{Fermi GBM Team}. 2024, GRB Coordinates Network, 37234, 1

\bibitem[{{Fern{\'a}ndez} \& {Davis}(2011)}]{fernandez2011}
{Fern{\'a}ndez}, R., \& {Davis}, S.~W. 2011, \apj, 730, 131,
  \dodoi{10.1088/0004-637X/730/2/131}

\bibitem[{{Fruscione} {et~al.}(2006){Fruscione}, {McDowell}, {Allen},
  {Brickhouse}, {Burke}, {Davis}, {Durham}, {Elvis}, {Galle}, {Harris},
  {Huenemoerder}, {Houck}, {Ishibashi}, {Karovska}, {Nicastro}, {Noble},
  {Nowak}, {Primini}, {Siemiginowska}, {Smith}, \&
  {Wise}}]{2006SPIE.6270E..1VF}
{Fruscione}, A., {McDowell}, J.~C., {Allen}, G.~E., {et~al.} 2006, in Society
  of Photo-Optical Instrumentation Engineers (SPIE) Conference Series, Vol.
  6270, Observatory Operations: Strategies, Processes, and Systems, ed. D.~R.
  {Silva} \& R.~E. {Doxsey}, 62701V, \dodoi{10.1117/12.671760}

\bibitem[{{GECAM team}(2024)}]{2024GCN.37240....1G}
{GECAM team}. 2024, GRB Coordinates Network, 37240, 1

\bibitem[{{Gnedin} \& {Pavlov}(1974)}]{1974JETP...38..903G}
{Gnedin}, Y.~N., \& {Pavlov}, G.~G. 1974, Soviet Journal of Experimental and
  Theoretical Physics, 38, 903

\bibitem[{{G{\"o}tz} {et~al.}(2006){G{\"o}tz}, {Mereghetti}, {Tiengo}, \&
  {Esposito}}]{2006A&A...449L..31G}
{G{\"o}tz}, D., {Mereghetti}, S., {Tiengo}, A., \& {Esposito}, P. 2006, \aap,
  449, L31, \dodoi{10.1051/0004-6361:20064870}

\bibitem[{{Harding} \& {Lai}(2006)}]{2006RPPh...69.2631H}
{Harding}, A.~K., \& {Lai}, D. 2006, Reports on Progress in Physics, 69, 2631,
  \dodoi{10.1088/0034-4885/69/9/R03}

\bibitem[{{Harrison} {et~al.}(2013){Harrison}, {Craig}, {Christensen},
  {Hailey}, {Zhang}, {Boggs}, {Stern}, {Cook}, {Forster}, {Giommi},
  {Grefenstette}, {Kim}, {Kitaguchi}, {Koglin}, {Madsen}, {Mao}, {Miyasaka},
  {Mori}, {Perri}, {Pivovaroff}, {Puccetti}, {Rana}, {Westergaard}, {Willis},
  {Zoglauer}, {An}, {Bachetti}, {Barri{\`e}re}, {Bellm}, {Bhalerao},
  {Brejnholt}, {Fuerst}, {Liebe}, {Markwardt}, {Nynka}, {Vogel}, {Walton},
  {Wik}, {Alexander}, {Cominsky}, {Hornschemeier}, {Hornstrup}, {Kaspi},
  {Madejski}, {Matt}, {Molendi}, {Smith}, {Tomsick}, {Ajello}, {Ballantyne},
  {Balokovi{\'c}}, {Barret}, {Bauer}, {Blandford}, {Brandt}, {Brenneman},
  {Chiang}, {Chakrabarty}, {Chenevez}, {Comastri}, {Dufour}, {Elvis}, {Fabian},
  {Farrah}, {Fryer}, {Gotthelf}, {Grindlay}, {Helfand}, {Krivonos}, {Meier},
  {Miller}, {Natalucci}, {Ogle}, {Ofek}, {Ptak}, {Reynolds}, {Rigby},
  {Tagliaferri}, {Thorsett}, {Treister}, \& {Urry}}]{2013ApJ...770..103H}
{Harrison}, F.~A., {Craig}, W.~W., {Christensen}, F.~E., {et~al.} 2013, \apj,
  770, 103, \dodoi{10.1088/0004-637X/770/2/103}

\bibitem[{{Heyl} {et~al.}(2024){Heyl}, {Taverna}, {Turolla}, {Israel}, {Ng},
  {K{\i}rm{\i}z{\i}bayrak}, {Gonz{\'a}lez-Caniulef}, {Caiazzo}, {Zane},
  {Ehlert}, {Negro}, {Agudo}, {Antonelli}, {Bachetti}, {Baldini},
  {Baumgartner}, {Bellazzini}, {Bianchi}, {Bongiorno}, {Bonino}, {Brez},
  {Bucciantini}, {Capitanio}, {Castellano}, {Cavazzuti}, {Chen}, {Ciprini},
  {Costa}, {De Rosa}, {Del Monte}, {Di Gesu}, {Di Lalla}, {Di Marco},
  {Donnarumma}, {Doroshenko}, {Dov{\v{c}}iak}, {Enoto}, {Evangelista},
  {Fabiani}, {Ferrazzoli}, {Garcia}, {Gunji}, {Hayashida}, {Iwakiri},
  {Jorstad}, {Kaaret}, {Karas}, {Kislat}, {Kitaguchi}, {Kolodziejczak},
  {Krawczynski}, {Monaca}, {Latronico}, {Liodakis}, {Maldera}, {Manfreda},
  {Marin}, {Marinucci}, {Marscher}, {Marshall}, {Massaro}, {Matt}, {Mitsuishi},
  {Mizuno}, {Muleri}, {Ng}, {O'Dell}, {Omodei}, {Oppedisano}, {Papitto},
  {Pavlov}, {Peirson}, {Perri}, {Pesce-Rollins}, {Petrucci}, {Pilia},
  {Possenti}, {Poutanen}, {Puccetti}, {Ramsey}, {Rankin}, {Ratheesh},
  {Roberts}, {Romani}, {Sgr{\`o}}, {Slane}, {Soffitta}, {Spandre}, {Swartz},
  {Tamagawa}, {Tavecchio}, {Tawara}, {Tennant}, {Thomas}, {Tombesi}, {Trois},
  {Tsygankov}, {Vink}, {Weisskopf}, {Wu}, \& {Xie}}]{2024MNRAS.52712219H}
{Heyl}, J., {Taverna}, R., {Turolla}, R., {et~al.} 2024, \mnras, 527, 12219,
  \dodoi{10.1093/mnras/stad3680}

\bibitem[{{Heyl} \& {Shaviv}(2002)}]{heyl2002}
{Heyl}, J.~S., \& {Shaviv}, N.~J. 2002, \prd, 66, 023002,
  \dodoi{10.1103/PhysRevD.66.023002}

\bibitem[{{Kaspi} \& {Beloborodov}(2017)}]{kaspi+belo17}
{Kaspi}, V.~M., \& {Beloborodov}, A.~M. 2017, \araa, 55, 261,
  \dodoi{10.1146/annurev-astro-081915-023329}

\bibitem[{{Kelly} {et~al.}(2024{\natexlab{a}}){Kelly}, {Gonz{\'a}lez-Caniulef},
  {Zane}, {Turolla}, \& {Taverna}}]{kelly+24}
{Kelly}, R. M.~E., {Gonz{\'a}lez-Caniulef}, D., {Zane}, S., {Turolla}, R., \&
  {Taverna}, R. 2024{\natexlab{a}}, \mnras, 534, 1355,
  \dodoi{10.1093/mnras/stae2163}

\bibitem[{{Kelly} {et~al.}(2024{\natexlab{b}}){Kelly}, {Zane}, {Turolla}, \&
  {Taverna}}]{2024MNRAS.528.3927K}
{Kelly}, R. M.~E., {Zane}, S., {Turolla}, R., \& {Taverna}, R.
  2024{\natexlab{b}}, \mnras, 528, 3927, \dodoi{10.1093/mnras/stae159}

\bibitem[{{Kuiper} {et~al.}(2006){Kuiper}, {Hermsen}, {den Hartog}, \&
  {Collmar}}]{2006ApJ...645..556K}
{Kuiper}, L., {Hermsen}, W., {den Hartog}, P.~R., \& {Collmar}, W. 2006, \apj,
  645, 556, \dodoi{10.1086/504317}

\bibitem[{{Kuiper} {et~al.}(2004){Kuiper}, {Hermsen}, \&
  {Mendez}}]{2004ApJ...613.1173K}
{Kuiper}, L., {Hermsen}, W., \& {Mendez}, M. 2004, \apj, 613, 1173,
  \dodoi{10.1086/423129}

\bibitem[{{Kumar} \& {Safi-Harb}(2010)}]{2010ApJ...725L.191K}
{Kumar}, H.~S., \& {Safi-Harb}, S. 2010, \apjl, 725, L191,
  \dodoi{10.1088/2041-8205/725/2/L191}

\bibitem[{{Kumar} {et~al.}(2014){Kumar}, {Safi-Harb}, {Slane}, \&
  {Gotthelf}}]{2014ApJ...781...41K}
{Kumar}, H.~S., {Safi-Harb}, S., {Slane}, P.~O., \& {Gotthelf}, E.~V. 2014,
  \apj, 781, 41, \dodoi{10.1088/0004-637X/781/1/41}

\bibitem[{{Laurent} {et~al.}(2021){Laurent}, {Acero}, {Beckmann}, {Brandt},
  {Cangemi}, {Civitani}, {Clavel}, {Coleiro}, {Curado}, {Ferrando}, {Ferrigno},
  {Frontera}, {Gastaldello}, {G{\"o}tz}, {Gouiff{\`e}s}, {Grinberg}, {Hanlon},
  {Hartmann}, {Maggi}, {Marin}, {Meuris}, {Okajima}, {Pareschi}, {Pratt},
  {Rea}, {Rodriguez}, {Rossetti}, {Spiga}, {Virgilli}, \&
  {Zane}}]{2021ExA....51.1143L}
{Laurent}, P., {Acero}, F., {Beckmann}, V., {et~al.} 2021, Experimental
  Astronomy, 51, 1143, \dodoi{10.1007/s10686-021-09723-x}

\bibitem[{{Mereghetti} {et~al.}(2005){Mereghetti}, {G{\"o}tz}, {Mirabel}, \&
  {Hurley}}]{2005A&A...433L...9M}
{Mereghetti}, S., {G{\"o}tz}, D., {Mirabel}, I.~F., \& {Hurley}, K. 2005, \aap,
  433, L9, \dodoi{10.1051/0004-6361:200500088}

\bibitem[{{Molkov} {et~al.}(2005){Molkov}, {Hurley}, {Sunyaev}, {Shtykovsky},
  {Revnivtsev}, \& {Kouveliotou}}]{2005A&A...433L..13M}
{Molkov}, S., {Hurley}, K., {Sunyaev}, R., {et~al.} 2005, \aap, 433, L13,
  \dodoi{10.1051/0004-6361:200500087}

\bibitem[{{Morii} {et~al.}(2003){Morii}, {Sato}, {Kataoka}, \&
  {Kawai}}]{2003PASJ...55L..45M}
{Morii}, M., {Sato}, R., {Kataoka}, J., \& {Kawai}, N. 2003, \pasj, 55, L45,
  \dodoi{10.1093/pasj/55.3.L45}

\bibitem[{{NICER Team}(2024)}]{2024ATel16789....1D}
{NICER Team}. 2024, The Astronomer's Telegram, 16789, 1

\bibitem[{{Olausen} \& {Kaspi}(2014)}]{2014ApJS..212....6O}
{Olausen}, S.~A., \& {Kaspi}, V.~M. 2014, \apjs, 212, 6,
  \dodoi{10.1088/0067-0049/212/1/6}

\bibitem[{{Radhakrishnan} \& {Cooke}(1969)}]{rvm}
{Radhakrishnan}, V., \& {Cooke}, D.~J. 1969, \aplett, 3, 225

\bibitem[{{Rybicki} \& {Lightman}(1979)}]{1979rpa..book.....R}
{Rybicki}, G.~B., \& {Lightman}, A.~P. 1979, {Radiative processes in
  astrophysics} (Wiley-Interscience)

\bibitem[{{Slane} {et~al.}(2024){Slane}, {Ferrazzoli}, {Zhou}, \&
  {Vink}}]{2024Galax..12...59S}
{Slane}, P., {Ferrazzoli}, R., {Zhou}, P., \& {Vink}, J. 2024, Galaxies, 12,
  59, \dodoi{10.3390/galaxies12050059}

\bibitem[{{Stewart} {et~al.}(2024){Stewart}, {Younes}, {Harding}, {Wadiasingh},
  {Baring}, {Negro}, {Strohmayer}, {Ho}, {Ng}, {Arzoumanian}, {Dinh Thi}, {Di
  Lalla}, {Enoto}, {Gendreau}, {Hu}, {van Kooten}, {Kouveliotou}, \&
  {McEwen}}]{2024arXiv241216036S}
{Stewart}, R., {Younes}, G., {Harding}, A., {et~al.} 2024, arXiv e-prints,
  arXiv:2412.16036, \dodoi{10.48550/arXiv.2412.16036}

\bibitem[{{Str{\"u}der} {et~al.}(2001){Str{\"u}der}, {Briel}, {Dennerl},
  {Hartmann}, {Kendziorra}, {Meidinger}, {Pfeffermann}, {Reppin}, {Aschenbach},
  {Bornemann}, {Br{\"a}uninger}, {Burkert}, {Elender}, {Freyberg}, {Haberl},
  {Hartner}, {Heuschmann}, {Hippmann}, {Kastelic}, {Kemmer}, {Kettenring},
  {Kink}, {Krause}, {M{\"u}ller}, {Oppitz}, {Pietsch}, {Popp}, {Predehl},
  {Read}, {Stephan}, {St{\"o}tter}, {Tr{\"u}mper}, {Holl}, {Kemmer}, {Soltau},
  {St{\"o}tter}, {Weber}, {Weichert}, {von Zanthier}, {Carathanassis}, {Lutz},
  {Richter}, {Solc}, {B{\"o}ttcher}, {Kuster}, {Staubert}, {Abbey}, {Holland},
  {Turner}, {Balasini}, {Bignami}, {La Palombara}, {Villa}, {Buttler},
  {Gianini}, {Lain{\'e}}, {Lumb}, \& {Dhez}}]{2001A&A...365L..18S}
{Str{\"u}der}, L., {Briel}, U., {Dennerl}, K., {et~al.} 2001, \aap, 365, L18,
  \dodoi{10.1051/0004-6361:20000066}

\bibitem[{{SVOM/GRM Team}(2024)}]{2024GCN.37297....1S}
{SVOM/GRM Team}. 2024, GRB Coordinates Network, 37297, 1

\bibitem[{{Swift Team}(2024{\natexlab{a}})}]{2024GCN.37211....1S}
{Swift Team}. 2024{\natexlab{a}}, GRB Coordinates Network, 37211, 1

\bibitem[{{Swift Team}(2024{\natexlab{b}})}]{2024GCN.37222....1S}
---. 2024{\natexlab{b}}, GRB Coordinates Network, 37222, 1

\bibitem[{{Taverna} {et~al.}(2014){Taverna}, {Muleri}, {Turolla}, {Soffitta},
  {Fabiani}, \& {Nobili}}]{taverna+14}
{Taverna}, R., {Muleri}, F., {Turolla}, R., {et~al.} 2014, \mnras, 438, 1686,
  \dodoi{10.1093/mnras/stt2310}

\bibitem[{{Taverna} \& {Turolla}(2024)}]{2024Galax..12....6T}
{Taverna}, R., \& {Turolla}, R. 2024, Galaxies, 12, 6,
  \dodoi{10.3390/galaxies12010006}

\bibitem[{{Taverna} {et~al.}(2015){Taverna}, {Turolla}, {Gonzalez Caniulef},
  {Zane}, {Muleri}, \& {Soffitta}}]{taverna+15}
{Taverna}, R., {Turolla}, R., {Gonzalez Caniulef}, D., {et~al.} 2015, \mnras,
  454, 3254, \dodoi{10.1093/mnras/stv2168}

\bibitem[{{Taverna} {et~al.}(2020){Taverna}, {Turolla}, {Suleimanov},
  {Potekhin}, \& {Zane}}]{taverna+20}
{Taverna}, R., {Turolla}, R., {Suleimanov}, V., {Potekhin}, A.~Y., \& {Zane},
  S. 2020, \mnras, 492, 5057, \dodoi{10.1093/mnras/staa204}

\bibitem[{{Taverna} {et~al.}(2022){Taverna}, {Turolla}, {Muleri}, {Heyl},
  {Zane}, {Baldini}, {Gonz{\'a}lez-Caniulef}, {Bachetti}, {Rankin}, {Caiazzo},
  {Di Lalla}, {Doroshenko}, {Errando}, {Gau}, {K{\i}rm{\i}z{\i}bayrak},
  {Krawczynski}, {Negro}, {Ng}, {Omodei}, {Possenti}, {Tamagawa}, {Uchiyama},
  {Weisskopf}, {Agudo}, {Antonelli}, {Baumgartner}, {Bellazzini}, {Bianchi},
  {Bongiorno}, {Bonino}, {Brez}, {Bucciantini}, {Capitanio}, {Castellano},
  {Cavazzuti}, {Ciprini}, {Costa}, {De Rosa}, {Del Monte}, {Di Gesu}, {Di
  Marco}, {Donnarumma}, {Dov{\v{c}}iak}, {Ehlert}, {Enoto}, {Evangelista},
  {Fabiani}, {Ferrazzoli}, {Garcia}, {Gunji}, {Hayashida}, {Iwakiri},
  {Jorstad}, {Karas}, {Kitaguchi}, {Kolodziejczak}, {La Monaca}, {Latronico},
  {Liodakis}, {Maldera}, {Manfreda}, {Marin}, {Marinucci}, {Marscher},
  {Marshall}, {Matt}, {Mitsuishi}, {Mizuno}, {Ng}, {O{\textquoteright}Dell},
  {Oppedisano}, {Papitto}, {Pavlov}, {Peirson}, {Perri}, {Pesce-Rollins},
  {Pilia}, {Poutanen}, {Puccetti}, {Ramsey}, {Ratheesh}, {Romani}, {Sgr{\`o}},
  {Slane}, {Soffitta}, {Spandre}, {Tavecchio}, {Tawara}, {Tennant}, {Thomas},
  {Tombesi}, {Trois}, {Tsygankov}, {Vink}, {Wu}, \& {Xie}}]{taverna+22}
{Taverna}, R., {Turolla}, R., {Muleri}, F., {et~al.} 2022, Science, 378, 646,
  \dodoi{10.1126/science.add0080}

\bibitem[{{Thompson} \& {Beloborodov}(2005)}]{2005ApJ...634..565T}
{Thompson}, C., \& {Beloborodov}, A.~M. 2005, \apj, 634, 565,
  \dodoi{10.1086/432245}

\bibitem[{{Thompson} \& {Duncan}(1993)}]{td93}
{Thompson}, C., \& {Duncan}, R.~C. 1993, \apj, 408, 194, \dodoi{10.1086/172580}

\bibitem[{{Tian} \& {Leahy}(2008)}]{2008ApJ...677..292T}
{Tian}, W.~W., \& {Leahy}, D.~A. 2008, \apj, 677, 292, \dodoi{10.1086/529120}

\bibitem[{{Turner} {et~al.}(2001){Turner}, {Abbey}, {Arnaud}, {Balasini},
  {Barbera}, {Belsole}, {Bennie}, {Bernard}, {Bignami}, {Boer}, {Briel},
  {Butler}, {Cara}, {Chabaud}, {Cole}, {Collura}, {Conte}, {Cros}, {Denby},
  {Dhez}, {Di Coco}, {Dowson}, {Ferrando}, {Ghizzardi}, {Gianotti}, {Goodall},
  {Gretton}, {Griffiths}, {Hainaut}, {Hochedez}, {Holland}, {Jourdain},
  {Kendziorra}, {Lagostina}, {Laine}, {La Palombara}, {Lortholary}, {Lumb},
  {Marty}, {Molendi}, {Pigot}, {Poindron}, {Pounds}, {Reeves}, {Reppin},
  {Rothenflug}, {Salvetat}, {Sauvageot}, {Schmitt}, {Sembay}, {Short},
  {Spragg}, {Stephen}, {Str{\"u}der}, {Tiengo}, {Trifoglio}, {Tr{\"u}mper},
  {Vercellone}, {Vigroux}, {Villa}, {Ward}, {Whitehead}, \&
  {Zonca}}]{2001A&A...365L..27T}
{Turner}, M.~J.~L., {Abbey}, A., {Arnaud}, M., {et~al.} 2001, \aap, 365, L27,
  \dodoi{10.1051/0004-6361:20000087}

\bibitem[{{Turolla} {et~al.}(2015){Turolla}, {Zane}, \& {Watts}}]{turolla+15}
{Turolla}, R., {Zane}, S., \& {Watts}, A.~L. 2015, Reports on Progress in
  Physics, 78, 116901, \dodoi{10.1088/0034-4885/78/11/116901}

\bibitem[{{Turolla} {et~al.}(2023){Turolla}, {Taverna}, {Israel}, {Muleri},
  {Zane}, {Bachetti}, {Heyl}, {Di Marco}, {Gau}, {Krawczynski}, {Ng},
  {Possenti}, {Poutanen}, {Baldini}, {Matt}, {Negro}, {Agudo}, {Antonelli},
  {Baumgartner}, {Bellazzini}, {Bianchi}, {Bongiorno}, {Bonino}, {Brez},
  {Bucciantini}, {Capitanio}, {Castellano}, {Cavazzuti}, {Chen}, {Ciprini},
  {Costa}, {De Rosa}, {Del Monte}, {Di Gesu}, {Di Lalla}, {Donnarumma},
  {Doroshenko}, {Dov{\v{c}}iak}, {Ehlert}, {Enoto}, {Evangelista}, {Fabiani},
  {Ferrazzoli}, {Garcia}, {Gunji}, {Hayashida}, {Iwakiri}, {Jorstad}, {Kaaret},
  {Karas}, {Kislat}, {Kitaguchi}, {Kolodziejczak}, {La Monaca}, {Latronico},
  {Liodakis}, {Maldera}, {Manfreda}, {Marin}, {Marinucci}, {Marscher},
  {Marshall}, {Massaro}, {Mitsuishi}, {Mizuno}, {Ng}, {O'Dell}, {Omodei},
  {Oppedisano}, {Papitto}, {Pavlov}, {Peirson}, {Perri}, {Pesce-Rollins},
  {Petrucci}, {Pilia}, {Puccetti}, {Ramsey}, {Rankin}, {Ratheesh}, {Roberts},
  {Romani}, {Sgr{\'o}}, {Slane}, {Soffitta}, {Spandre}, {Swartz}, {Tamagawa},
  {Tavecchio}, {Tawara}, {Tennant}, {Thomas}, {Tombesi}, {Trois}, {Tsygankov},
  {Vink}, {Weisskopf}, {Wu}, \& {Xie}}]{2023ApJ...954...88T}
{Turolla}, R., {Taverna}, R., {Israel}, G.~L., {et~al.} 2023, \apj, 954, 88,
  \dodoi{10.3847/1538-4357/aced05}

\bibitem[{{Vasisht} \& {Gotthelf}(1997)}]{1997ApJ...486L.129V}
{Vasisht}, G., \& {Gotthelf}, E.~V. 1997, \apjl, 486, L129,
  \dodoi{10.1086/310843}

\bibitem[{{Vogel} {et~al.}(2014){Vogel}, {Hasco{\"e}t}, {Kaspi}, {An},
  {Archibald}, {Beloborodov}, {Boggs}, {Christensen}, {Craig}, {Gotthelf},
  {Grefenstette}, {Hailey}, {Harrison}, {Kennea}, {Madsen}, {Pivovaroff},
  {Stern}, \& {Zhang}}]{2014ApJ...789...75V}
{Vogel}, J.~K., {Hasco{\"e}t}, R., {Kaspi}, V.~M., {et~al.} 2014, \apj, 789,
  75, \dodoi{10.1088/0004-637X/789/1/75}

\bibitem[{{Wachter} {et~al.}(2004){Wachter}, {Patel}, {Kouveliotou}, {Bouchet},
  {{\"O}zel}, {Tennant}, {Woods}, {Hurley}, {Becker}, \&
  {Slane}}]{2004ApJ...615..887W}
{Wachter}, S., {Patel}, S.~K., {Kouveliotou}, C., {et~al.} 2004, \apj, 615,
  887, \dodoi{10.1086/424704}

\bibitem[{{Wadiasingh} {et~al.}(2018){Wadiasingh}, {Baring}, {Gonthier}, \&
  {Harding}}]{2018ApJ...854...98W}
{Wadiasingh}, Z., {Baring}, M.~G., {Gonthier}, P.~L., \& {Harding}, A.~K. 2018,
  \apj, 854, 98, \dodoi{10.3847/1538-4357/aaa460}

\bibitem[{{Weisskopf} {et~al.}(2010){Weisskopf}, {Guainazzi}, {Jahoda},
  {Shaposhnikov}, {O'Dell}, {Zavlin}, {Wilson-Hodge}, \&
  {Elsner}}]{Weisskopf2010}
{Weisskopf}, M.~C., {Guainazzi}, M., {Jahoda}, K., {et~al.} 2010, \apj, 713,
  912, \dodoi{10.1088/0004-637X/713/2/912}

\bibitem[{Weisskopf {et~al.}(2022)Weisskopf, Soffitta, Baldini, Ramsey, O'Dell,
  Romani, Matt, Deininger, Baumgartner, Bellazzini, Costa, Kolodziejczak,
  Latronico, Marshall, Muleri, Bongiorno, Tennant, Bucciantini, Dovciak, Marin,
  Marscher, Poutanen, Slane, Turolla, Kalinowski, {Di Marco}, Fabiani, Minuti,
  Monaca, Pinchera, Rankin, Sgrò, Trois, Xie, Alexander, Allen, Amici,
  Andersen, Antonelli, Antoniak, Attiná, Barbanera, Bachetti, Baggett, Bladt,
  Brez, Bonino, Boree, Borotto, Breeding, Brienza, Bygott, Caporale, Cardelli,
  Carpentiero, Castellano, Castronuovo, Cavalli, Cavazzuti, Ceccanti, Centrone,
  Citraro, D'Amico, D'Alba, Gesu, Monte, Dietz, Lalla, Persio, Dolan,
  Donnarumma, Evangelista, Ferrant, Ferrazzoli, Ferrie, Footdale, Forsyth,
  Foster, Garelick, Gunji, Gurnee, Head, Hibbard, Johnson, Kelly, Kilaru,
  Lefevre, Roy, Loffredo, Lorenzi, Lucchesi, Maddox, Magazzu, Maldera,
  Manfreda, Mangraviti, Marengo, Marrocchesi, Massaro, Mauger, McCracken,
  McEachen, Mize, Mereu, Mitchell, Mitsuishi, Morbidini, Mosti, Nasimi, Negri,
  Negro, Nguyen, Nitschke, Nuti, Onizuka, Oppedisano, Orsini, Osborne, Pacheco,
  Paggi, Painter, Pavelitz, Pentz, Piazzolla, Perri, Pesce-Rollins, Peterson,
  Pilia, Profeti, Puccetti, Ranganathan, Ratheesh, Reedy, Root, Rubini,
  Ruswick, Sanchez, Sarra, Santoli, Scalise, Sciortino, Schroeder, Seek,
  Sosdian, Spandre, Speegle, Tamagawa, Tardiola, Tobia, Thomas, Valerie,
  Vimercati, Walden, Weddendorf, Wedmore, Welch, Zanetti, \&
  Zanetti}]{Weisskopf2022}
Weisskopf, M.~C., Soffitta, P., Baldini, L., {et~al.} 2022, Journal of
  Astronomical Telescopes, Instruments, and Systems, 8, 1 ,
  \dodoi{10.1117/1.JATIS.8.2.026002}

\bibitem[{{Yang} {et~al.}(2016){Yang}, {Archibald}, {Vogel}, {An}, {Kaspi},
  {Guillot}, {Beloborodov}, \& {Pivovaroff}}]{2016ApJ...831...80Y}
{Yang}, C., {Archibald}, R.~F., {Vogel}, J.~K., {et~al.} 2016, \apj, 831, 80,
  \dodoi{10.3847/0004-637X/831/1/80}

\bibitem[{{Younes} {et~al.}(2024){Younes}, {Hu}, {Enoto}, {Gendreau},
  {Arzoumanian}, {Hare}, {Ng}, \& {Wadiasingh}}]{2024ATel16802....1Y}
{Younes}, G., {Hu}, C.~P., {Enoto}, T., {et~al.} 2024, The Astronomer's
  Telegram, 16802, 1

\bibitem[{{Zane} {et~al.}(2023){Zane}, {Taverna}, {Gonz{\'a}lez-Caniulef},
  {Muleri}, {Turolla}, {Heyl}, {Uchiyama}, {Ng}, {Tamagawa}, {Caiazzo}, {Di
  Lalla}, {Marshall}, {Bachetti}, {La Monaca}, {Gau}, {Di Marco}, {Baldini},
  {Negro}, {Omodei}, {Rankin}, {Matt}, {Pavlov}, {Kitaguchi}, {Krawczynski},
  {Kislat}, {Kelly}, {Agudo}, {Antonelli}, {Baumgartner}, {Bellazzini},
  {Bianchi}, {Bongiorno}, {Bonino}, {Brez}, {Bucciantini}, {Capitanio},
  {Castellano}, {Cavazzuti}, {Chen}, {Ciprini}, {Costa}, {De Rosa}, {Del
  Monte}, {Di Gesu}, {Donnarumma}, {Doroshenko}, {Dov{\v{c}}iak}, {Ehlert},
  {Enoto}, {Evangelista}, {Fabiani}, {Ferrazzoli}, {Garcia}, {Gunji},
  {Hayashida}, {Iwakiri}, {Jorstad}, {Kaaret}, {Karas}, {Kolodziejczak},
  {Latronico}, {Liodakis}, {Maldera}, {Manfreda}, {Marin}, {Marinucci},
  {Marscher}, {Massaro}, {Mitsuishi}, {Mizuno}, {Ng}, {O'Dell}, {Oppedisano},
  {Papitto}, {Peirson}, {Perri}, {Pesce-Rollins}, {Petrucci}, {Pilia},
  {Possenti}, {Poutanen}, {Puccetti}, {Ramsey}, {Ratheesh}, {Roberts},
  {Romani}, {Sgr{\'o}}, {Slane}, {Soffitta}, {Spandre}, {Swartz}, {Tavecchio},
  {Tawara}, {Tennant}, {Thomas}, {Tombesi}, {Trois}, {Tsygankov}, {Vink},
  {Weisskopf}, {Wu}, \& {Xie}}]{zane+23}
{Zane}, S., {Taverna}, R., {Gonz{\'a}lez-Caniulef}, D., {et~al.} 2023, \apjl,
  944, L27, \dodoi{10.3847/2041-8213/acb703}

\end{thebibliography}
\bibliographystyle{aasjournal}



\end{document}